\documentclass[prd,reprint,amsmath,aps,superscriptaddress,nofootinbib]{revtex4-1}
\usepackage{geometry}
\usepackage{titlesec}
\usepackage{bm}
\usepackage{gensymb}
 \usepackage{graphicx}
 \usepackage{tensor}
 \usepackage{verbatim}
 \usepackage{courier}
 \usepackage{mathtools}
  \usepackage[inline]{enumitem}
 \usepackage{calc}
\usepackage{color}
\usepackage{amsmath}
\usepackage{xfrac}

\begin{document}
\title{Two-harmonic approximation for gravitational waveforms from precessing binaries}
\author{Stephen Fairhurst}
\affiliation{School of Physics and Astronomy, Cardiff University, Cardiff, CF24 3AA, United Kingdom}
\author{Rhys Green}
\affiliation{School of Physics and Astronomy, Cardiff University, Cardiff, CF24 3AA, United Kingdom}
\author{Charlie Hoy}
\affiliation{School of Physics and Astronomy, Cardiff University, Cardiff, CF24 3AA, United Kingdom}
\author{Mark Hannam}
\affiliation{School of Physics and Astronomy, Cardiff University, Cardiff, CF24 3AA, United Kingdom}
\author{Alistair Muir}
\affiliation{School of Physics and Astronomy, Cardiff University, Cardiff, CF24 3AA, United Kingdom}
\date{\today}

\begin{abstract}
Binary-black-hole orbits precess when the black-hole spins are mis-aligned with the binary's orbital angular
momentum. The apparently complicated dynamics can in most cases be described as simple precession of
the orbital angular momentum about an approximately fixed total angular momentum. However, the imprint of
the precession on the observed gravitational-wave signal is yet more complicated, with a non-trivial time-varying
dependence on  black-hole dynamics, the binary's orientation and the detector polarization. As a result,
it is difficult to predict under which conditions precession effects are measurable in
gravitational-wave observations, and their impact on both signal detection and source characterization.
We show that the observed waveform can be simplified by decomposing it as a power series in a new
precession parameter $b = \tan(\beta/2)$, where $\beta$ is the opening angle between the orbital and total
angular momenta. The power series is made up of five harmonics, with frequencies that differ by the
binary's precession frequency, and individually do not exhibit amplitude and phase modulations.  
In many cases, the waveform can be well approximated by the
two leading harmonics.  In this approximation we are able to obtain a simple picture of precession as caused 
by the beating of two waveforms of similar frequency.  This enables us to identify regions of the parameter
space where precession is likely to have an observable effect on the waveform, and to propose a new approach
to searching for signals from precessing binaries, based upon the two-harmonic approximation.
\end{abstract}

\maketitle

\section{Introduction}
\label{sec:intro}

When the spins of black holes in a binary system are mis-aligned with the binary's orbital angular
momentum, both the spins and orbital angular momentum will precess \cite{Apostolatos:1994mx,Kidder:1995zr,
Apostolatos:1995pj, Buonanno:2002fy}.  We therefore expect that
most astrophysical binaries will undergo precession, but to date there has been no evidence
of precession in gravitational-wave (GW) observations from the Advanced LIGO and Virgo detectors
\cite{LIGOScientific:2018mvr, LIGOScientific:2018jsj}.
This is not necessarily surprising, because precession often leaves only a weak imprint on the observable signal,
particularly when the black holes are of comparable mass and the binary's orbit is face-on to the detector,
which are the most likely configurations that have been observed so far. Despite this heuristic picture, there
is no simple means to estimate the measurability of precession of a given binary configuration, and as such it
is difficult to predict when precession effects will be conclusively observed in GW events.

Detailed parameter estimation techniques have been developed, which enable the reconstruction
of the parameters of observed signals \cite{vanderSluys:2008qx, Veitch:2009hd, Veitch:2014wba,
Biwer:2018osg, Smith:2016qas}, in addition to approximate Fisher-matrix methods~\cite{Poisson:1995ef,Lang:1900bz}. 
In parallel,
techniques have been developed that provide an intuitive understanding
of the measurement accuracy of certain parameters (or parameter 
combinations)~ \cite{Fairhurst:2009tc, Baird:2012cu, Hannam:2013uu,
Singer:2015ema, Vallisneri:2012qq, Usman:2018imj}.
These have typically involved either approximations (such as leading order, Fisher Matrix type calculations),
restriction to a subset of system parameters (for example masses and spins; timing and
sky location; binary orientation).  Combined, these give an understanding of the accuracy
of parameter estimation for non-precessing systems.  

In parallel, there have been significant developments in understanding the implications of precession, starting
with the early work in Refs.~\cite{Apostolatos:1994mx,Apostolatos:1995pj,  Buonanno:2002fy} which provided
insights into the impact of precession on the gravitational waveform emitted during the inspiral of compact binaries.
Subsequently, black hole binary waveforms which incorporate precession through merger have been developed
\cite{Hannam:2013oca, Taracchini:2013rva, Khan:2019kot, Varma:2019csw, Pratten:2020ceb}; large scale parameter 
estimation studies of precession have 
been performed to identify the regions of parameter space
where precession will be observable \cite{Vitale:2014mka, Trifiro:2015zda, Vitale:2016avz, OShaughnessy:2014shr, 
Farr:2014qka, Littenberg:2016dgj, Farr:2015lna, Vitale:2016icu}; and new theoretical insights into the impact of precession 
on both detection and parameter estimation have been obtained \cite{Brown:2012gs,o2020semianalytic, Lundgren:2013jla}.
Complementary to this, there have been several efforts to understand the impact of precession on searches 
\cite{Harry:2013tca, Brown:2012gs}, and to implement searches for precessing signals 
\cite{Apostolatos:1995pj, Fazi:2009ifa, Harry:2011qh, Harry:2016ijz, Buonanno:2002fy}.  This has led to an increasingly
clear picture of the impact of precession: it is most significant for binaries with large mass ratios, where the in-plane
spin components are large and for systems where the total angular momentum is mis-aligned with the line of sight.

At leading order, the gravitational waveform emitted by a precessing binary is composed of five harmonics, which
are offset by multiples of the precession frequency \cite{Hannam:2013oca, Lundgren:2013jla}.  We
show that these harmonics form a natural hierarchy with the amplitude of the sub-leading harmonics suppressed
by a factor that depends upon the opening angle (the angle between the orbital
and total angular momenta).  Using this approximation, and restricting to the two
leading harmonics, we are able to obtain relatively simple expressions for the
precession waveform.  Each harmonic takes the form of a non-precessing-binary
waveform (i.e., with monotonic amplitude and frequency evolution during the inspiral of non-eccentric systems),
and the amplitude and phase modulations of the complete precessing-binary waveform
arise as beating between the two harmonics.

The purpose of this paper is to introduce this decomposition (Sec.~\ref{sec:waveform}), with
an alternative derivation given in the Appendix,
and the two-harmonic approximation (Sec.~\ref{sec:twoharmonics}),
and to identify its range of validity and accuracy (Sec.~\ref{sec:validity}). We then  
discuss a proposed search for precessing binaries
using the two-harmonic approximation (Sec.~\ref{sec:prec_search}) and finally introduce the
notion of a ``precession SNR'' that can be used to determine whether precession effects are observable
in a given system (Sec.~\ref{sec:obs_prec}).
We begin in the next section with a summary of precession in black-hole binaries.

\section{Black hole Spin Induced Precession}
\label{sec:precession}

In the general theory of relativity a binary consisting of two objects of masses, $m_{1}$ and $m_{2}$ (where we choose
$m_1 \ge m_2$ and denote $q = m_1/m_2$, so that $q \ge 1$), with spin angular momenta
$\mathbf{S_{1}}$ and $\mathbf{S_{2}}$, orbiting each other with angular momentum $\mathbf{L}$, will slowly inspiral due to the
loss of energy and momentum through the emission of gravitational waves.
If $\mathbf{S}_{1}\parallel\mathbf{S}_{2}\parallel\mathbf{L}$,
then the plane of the orbit remains fixed and in non-eccentric binaries the amplitude and frequency of the emitted
gravitational wave increases as the orbital separation decreases. The system eventually merges and forms a single
perturbed black hole that emits gravitational radiation as a superposition of quasinormal ringdown multipoles, until the
system settles down to its final state \cite{Kokkotas1999}.
\par
For the case where the total spin is not aligned with the total orbital angular momentum,
$\left(\mathbf{S}_{1}+\mathbf{S}_{2}\right)\times \mathbf{L} \neq 0$, in most cases the orbital plane of the binary will precess around the approximately constant total angular momentum $\mathbf{J} = \mathbf{S_{1}} + \mathbf{S_{2}} + \mathbf{L}$, i.e., $\mathbf{L}$ precesses
around $\mathbf{J}$, and the spins precess such that $\dot{\mathbf{S}} = - \dot{\mathbf{L}}$~\cite{Apostolatos:1994mx}. For configurations where
$\mathbf{J} \approx 0$, the system undergoes ``transitional precession''~\cite{Apostolatos:1994mx,Kidder:1995zr}, but this is 
expected to be rare in LIGO-Virgo detections.
The angle between $\mathbf{L}$ and $\mathbf{J}$ is
denoted by $\beta$. In simple precession cases $\beta$ slowly increases during inspiral as $L$ decreases
(recall that in the Newtonian limit $L \propto \sqrt{r}$, where $r$ is the orbital separation), but the spin magnitudes
$S_1$ and $S_2$ remain fixed, and, to a good approximation, so do their orbit-averaged components parallel and perpendicular to
$L$, $S_{i ||}$ and $S_{i \perp}$. The opening angle $\beta$ typically varies very little over the portion of a binary's inspiral that is
visible in a GW detector, and so it is often possible to make the approximation that $\beta$ is constant. This approximation has been
used to good effect in Ref.~\cite{Brown:2012gs}, and we will also use it in some of the discussion in this paper.

\begin{figure}[t]
	\centering
	\includegraphics[scale=0.25]{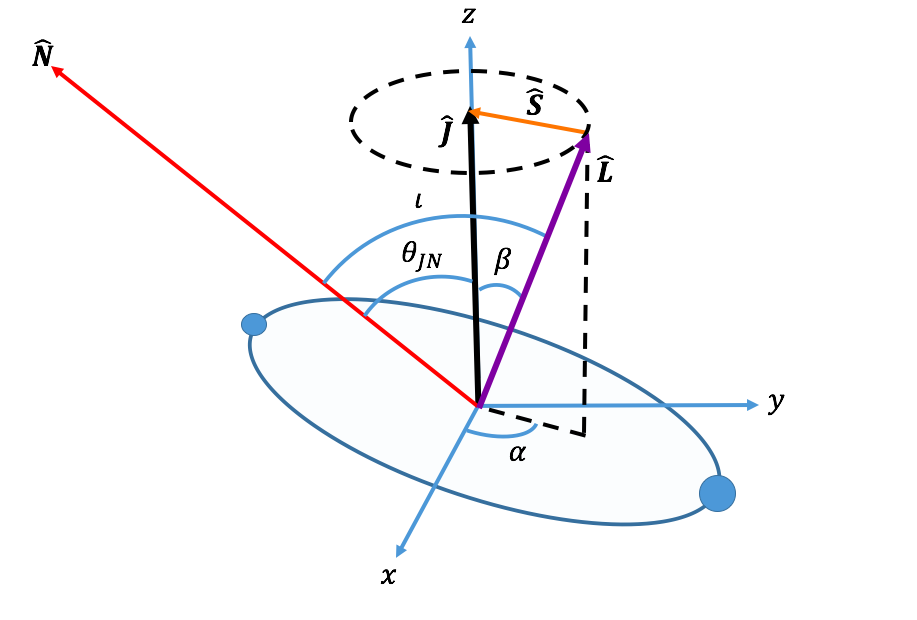}
	\caption{Plot showing how the precession angles used in this study are defined in the $J$-aligned frame. The normal vector here indicates the line of sight of the observer, $\mathbf{\hat{L}}$ and $\mathbf{\hat{J}}$ are the orbital angular momentum and total angular momentum vectors respectively, $S_{1x}, S_{1y}$ and $S_{1z}$ are the x, y and z components of the spin on the larger black hole.
	}
	\label{fig:precession_angles}
\end{figure}

Adopting the notation that the inclination angle of the binary as seen by an observer, $\iota$, is the angle between the orbital
angular momentum and the line of sight (see Fig.\ref{fig:precession_angles}), $\cos\iota = \mathbf{\hat{L}}\cdot\mathbf{\hat{N}}$,
where a caret denotes a unit vector
(e.g. $\hat{\mathbf{a}}=\mathbf{a}/|\mathbf{a}|$), the binary's orbital inclination becomes times
dependent. As a result the energy emitted in GWs in the $\mathbf{\hat{N}}$ direction will also be time dependent, where the maximum
instantaneous energy emission is approximately in the direction of $\mathbf{\hat{L}}$. If $\mathbf{\hat{N}}$ is aligned with $\mathbf{\hat{J}}$,
then $\iota \approx \beta$ and varies slowly and with minimal oscillations due only to orbital nutation. If $\mathbf{\hat{N}}$ is in
some other direction, then the energy emission will be modulated on the precession timescale. In the following we will not use
the inclination $\iota$, but rather combinations of $\beta$ and the angle between $\mathbf{J}$ and $\mathbf{\hat{N}}$, denoted by
$\theta_{\rm JN}$. As noted previously, $\mathbf{\hat{J}}$ is approximately constant
for simple precession cases, and we will treat it as a constant in the analysis in Sec.~\ref{sec:waveform}.

The signal measured in a detector will exhibit modulations in phase and amplitude that depend on $\beta$, $\theta_{\rm JN}$, the precession
angle of $\mathbf{L}$ around $\mathbf{J}$, denoted by $\alpha$, and the polarisation $\psi$ of the observed signal.
These angles are illustrated in
Fig.~\ref{fig:precession_angles}, and discussed further in Sec.~\ref{sec:waveform}. For now we note several well-known features of
precession waveforms~\cite{Apostolatos:1994mx,Kidder:1995zr}, which will be further sharpened in the discussion later in the paper.
The strength of precession in a system is
characterised by the degree of tilt of the binary's orbit, given by $\beta$, and by the precession frequency $\Omega_{P}$ of 
$\mathbf{L}$ around $\mathbf{J}$, which is given by
\begin{equation}
\Omega_p = \dot{\alpha} \, .
\end{equation}
 The angle $\beta$ is determined primarily by the total spin in the plane, and binary's mass ratio
and separation.
At leading order we can write the orbital angular momentum of the system as $L = \mu \sqrt{M r}$, where $\mu$ is the reduced mass,
$\mu = m_1 m_2 / M = qM/(1+q)^{2}$, and so to first approximation,
\begin{equation} \label{eq:opening_angle}
	\tan \beta = \frac{S_{\perp}}{\mu\sqrt{M r} + S_{\parallel}},
\end{equation}
which provides us with the basic dependence of $\beta$ on the binary configuration. At leading order the precession
frequency can be written as,
\begin{equation} \label{angularvelocity}
	\Omega_{p} \approx \left(2+\frac{3}{2q}\right)\frac{J}{r^{3}},
\end{equation} 
meaning that to first approximation it {\it does not} depend on the spins (or therefore the opening angle $\beta$), but
only on the binary's total mass, mass-ratio, and separation (or equivalently orbital frequency). The number of
precession cycles over a certain time or frequency range (e.g., over the course of an observation), depends
on the total mass and mass-ratio of the binary. In a GW observation there is a partial degeneracy between the mass ratio 
and the aligned spin
$S_{||}$~\cite{Cutler:1994ys,Poisson:1995ef,Baird:2012cu}, 
meaning that one of the chief effects of a measurement of precession will be to improve the
measurement of these two physical properties~\cite{OShaughnessy:2014shr}.

In the remainder of this paper we choose to describe the gravitational wave signal, precessing or non-precessing, with the 
{\tt IMRPhenomPv2} phenomenological model presented in Ref.~\cite{Hannam:2013oca}. This model exploits the phenomenology 
of simply precessing binaries described earlier, with the additional approximation that a precessing-binary waveform can be 
factorised into an underlying non-precessing waveform, and the precessional dynamics~\cite{Schmidt:2012rh}. 
The underlying
non-precessing-binary model is {\tt IMRPhenomD}~\cite{Husa:2015iqa, Khan:2015jqa}, using
only the spin components aligned with $\mathbf{L}$. In {\tt IMRPhenomD} both aligned spin components are used to generate an
approximate post-Newtonian phasing and amplitude, with corrections provided by fits to numerical-relativity waveforms, that are
parameterised by two different combinations of the two spin components. Although {\tt IMRPhenomD} has been found to model
well two-spin systems~\cite{Kumar:2016dhh}, its dominant spin dependence can be characterised well by the effective spin,
\begin{equation} \label{chi_eff}
	\chi_{\text{eff}} = \frac{1}{M} \left(\frac{\mathbf{S_{1}}}{m_{1}} + \frac{\mathbf{S_{2}}}{m_{2}}\right)\cdot \hat{\mathbf{L}},
\end{equation}
which takes values between $-1$ (both maximal anti-aligned spins) and $+1$ (both maximal aligned spins) to
describe the magnitude of spin aligned with the total angular momentum. For a given configuration {\tt IMRPhenomPv2} uses
the corresponding {\tt IMRPhenomD} waveform, but with the
final spin modified to take into account the in-plane spin components. A frequency-dependent rotation is then applied to the
non-precessing waveform to introduce the precession dynamics, which are modelled by frequency-domain post-Newonian
expressions for the precession angles for an approximately equivalent single-spin
system~\cite{Schmidt:2014iyl, Hannam:2013oca}, where the large black hole has spin,
\begin{equation} \label{chi_p}
	\chi_{\text{p}} = \frac{1}{A_{1} m_{1}^{2} }\text{max}\left(A_{1}S_{1\perp},A_{2}S_{2\perp}\right),
\end{equation}
where $A_{1} = 2+3q/2$ and $A_{2} = 2+3/(2q)$ and $\mathbf{S}_{i\perp}$ is the component of the spin perpendicular to $\mathbf{L}$. The effective precession spin parameter is obtained by averaging the relative in-plane spin orientation over a precession cycle, and so more accurate for a system that undergoes many precession cycles.

There are several important features which are not incorporated in the {\tt IMRPhenomPv2} waveform.  These include 
two-spin effects \cite{Taracchini:2013rva, Babak:2016tgq, Khan:2018fmp}, gravitational wave multipoles other than the leading 
22 mode \cite{Khan:2019kot}, significant precession during merger \cite{OShaughnessy:2012iol}, and spin alignment due to 
spin-orbit resonances during inspiral \cite{Gerosa:2014kta, Gerosa:2015tea}.  Some of these effects will have an impact
upon the distributions of black hole spin orientations when the binaries enter the LIGO or Virgo sensitivity band while
others can leave imprints on the waveform which may be observable, particularly close to the merger.   Nonetheless, the 
{\tt IMRPhenomPv2} has been used in the analysis of all LIGO-Virgo observations during the first two observing runs
\cite{Abbott:2016blz, TheLIGOScientific:2016pea, LIGOScientific:2018jsj, LIGOScientific:2018mvr}, and it captures much of
the dominant phenomenology of precessing-binary waveforms.  In addition, the decomposition presented in the next section
is in no way tied to the particular waveform used and could be equally well applied to other waveform models for precessing 
binaries which, for example, incorporate two-spin effects and precession during merger.  The current formalism does not 
include additional gravitational wave multipoles, and we will investigate this in a future work.  We expect the broad features
of many of the results presented in the remainder of the paper to be relatively unaffected by the specific waveform choice, but
the details for any specific signal could change.

\section{Harmonic decomposition of the waveform from a precessing binary}
 \label{sec:waveform}

The gravitational waveform emitted by a precessing system, as observed at a gravitational
wave detector, can be expressed approximately as \cite{Buonanno:2002fy, Brown:2012gs}
\begin{eqnarray}
	h(t) &=&
	- \left(\frac{d_{o}}{d_{L}}\right) A_{o}(t) \, \mathrm{Re} \Big[ e^{2 i \Phi_{S}(t)}  \nonumber \\
	&& \qquad \big(F_{+} (C_{+} - i S_{+}) + F_{\times} (C_{\times} - i S_{\times}) \big) \Big] \, .
	\label{eq:prec_wf}
\end{eqnarray}
Here, $A_{o}(t)$ denotes the amplitude of the gravitational wave signal in a (time-varying) frame aligned
with the orbital angular momentum of the binary, and depends upon the masses and spins of the binary.
Since the amplitude scales linearly with the luminosity distance,
we have chosen to introduce a fiducial normalization $A_{o}(t)$ for a waveform at a distance $d_{o}$ and
explicitly extract the distance dependence.%
\footnote{Of course, the observed waveform is also affected by the redshifting of frequencies.  For the calculation
discussed here, we work in the detector frame and consider the \textit{observed} masses, which are $(1+z)$ times
the source frame masses.}
$\Phi_{S}$ is the phase evolution in the
frame aligned with the total angular momentum $\mathbf{J}$ of the binary.  The phase evolution, $\Phi_{S}$,
is related to the orbital phase, $\phi_{orb}$, as
\begin{equation}\label{eq:phi_s}
	\Phi_{S}(t) = \phi_{orb}(t) - \epsilon (t)
\end{equation}
where \cite{Boyle:2011gg}
\begin{equation}
	\dot{\epsilon}(t) := \dot{\alpha}(t) \cos{\beta}(t)
\end{equation}
and, as before, $\beta$ is the opening angle and $\alpha$ gives the phase of the precession
of $\mathbf{L}$ around $\mathbf{J}$ as shown in Fig.~\ref{fig:precession_angles}. 
$F_{+}$ and $F_{\times}$ give
 the detector response relative to the $\mathbf{J}$-aligned frame and $C_{+, \times}$, $S_{+, \times}$
encode the time-varying response to the gravitational wave due to the evolution of the binary's
orbit relative to the detector.   They depend upon the three angles introduced previously:
 the precession opening angle $\beta$ and phase $\alpha$ and the angle between the total orbital
angular momentum and the line of sight $\theta_{\rm JN}$.
In terms of these angles, we can express $C_{+, \times}$ and $S_{+, \times}$ as%
\footnote{We have re-written the $C_{+}$ term relative to what is normally given in the literature, e.g. \cite{Buonanno:2002fy,
Brown:2012gs}, to group terms with the same $\alpha$ dependence.}
 \begin{eqnarray}
C_{+} &=& - \left(\frac{1 + \cos^{2} \theta_{\rm JN}}{2} \right) \left(\frac{1 + \cos^{2} \beta}{2} \right) \cos 2\alpha
	\nonumber \\
	&&- \frac{\sin 2\theta_{\rm JN}}{2} \frac{\sin 2\beta}{2} \cos\alpha
	- \frac{3}{4} \sin^{2}\theta_{\rm JN} \sin^{2}\beta, \nonumber \\
S_{+} &=& \left( \frac{1 + \cos^{2} \theta_{\rm JN}}{2}\right) \cos \beta \sin 2\alpha
	+ \frac{\sin 2\theta_{\rm JN}}{2} \sin \beta \sin \alpha, \nonumber \\
C_{\times} &=& - \cos \theta_{\rm JN} \left(\frac{1 + \cos^2 \beta}{2}\right) \sin 2 \alpha
	- \sin \theta_{\rm JN} \frac{\sin 2\beta}{2} \sin\alpha, \nonumber \\
S_{\times} &=& - \cos\theta_{\rm JN} \cos\beta \cos 2\alpha
	- \sin \theta_{\rm JN} \sin \beta \cos \alpha. \label{eq:cs_plus_cross}
\end{eqnarray}
The non-precessing expressions can be recovered in the limit of $\beta \rightarrow 0$ and
$\alpha \rightarrow \mathrm{constant}$ (which is then degenerate with the polarization of the system).
When $\beta$ is non-zero, the effect of precession is to modulate the detector response at
frequencies $\Omega_{P}$ and $2 \Omega_{P}$.  To make the harmonic content of $C_{+, \times}$
and $S_{+, \times}$ more explicit, we first introduce the parameter,
\begin{equation}
	b = \tan{(\beta/2)} \, ,
\end{equation}
and write the response functions in terms of it.  The terms involving $\beta$ can be expressed as
\begin{eqnarray}
\frac{1 + \cos^{2} \beta}{2} &=& \frac{1 + b^{4}}{(1 + b^{2})^{2}}, \nonumber \\
\cos \beta &=& \frac{1 - b^{4}}{(1 + b^{2})^{2}}, \nonumber \\
\frac{\sin 2\beta}{2} &=& \frac{2b (1 - b^{2})}{(1 + b^{2})^{2}}, \nonumber \\
\sin \beta &=& \frac{2b (1 + b^{2})}{(1 + b^{2})^{2}}, \nonumber \\
\sin^{2} \beta &=& \frac{4b^{2}}{(1 + b^{2})^{2}} \, .
\label{eq:trig_b}
\end{eqnarray}
Substituting the trigonometric identities from Eq.~(\ref{eq:trig_b}) into the expressions for $C_{+}$
and $S_{+}$ in Eq.~(\ref{eq:cs_plus_cross}) we obtain,
\begin{eqnarray}
\left(\frac{d_{o}}{d_{L}}\right) \left(C_{+} - i S_{+}\right)
	&=& - e^{2i\alpha} \sum_{k=0}^{4} \mathcal{A}^{+}_{k} \left[\frac{b^k e^{-ik\alpha}}{(1 + b^{2})^{2}}\right], \nonumber \\
\left(\frac{d_{o}}{d_{L}}\right) \left(C_{\times} - i S_{\times} \right)
	&=& i e^{2i\alpha} \sum_{k=0}^{4} \mathcal{A}^{\times}_{k} \left[\frac{b^k e^{-ik\alpha}}{(1 + b^{2})^{2}}\right],
\label{eq:cs_harms}
\end{eqnarray}
where we have introduced $\mathcal{A}_{k}^+$ and $\mathcal{A}_{k}^{\times}$ as
\begin{align}
\mathcal{A}_{0}^+ = \mathcal{A}_{4}^+ &= \frac{d_{o}}{d_L} \left(\frac{1+\cos^2 \theta_{\rm JN}}{2} \right), \nonumber\\
\mathcal{A}_{0}^{\times} = - \mathcal{A}_{4}^{\times} &= \frac{d_{o}}{d_L}\cos \theta_{\rm JN}, \nonumber \\
\mathcal{A}_{1}^+ = - \mathcal{A}_{3}^+ &= 2 \frac{d_{o}}{d_L} \sin\theta_{\rm JN} \cos\theta_{\rm JN}, \nonumber\\
\mathcal{A}_{1}^{\times} = \mathcal{A}_{3}^{\times} &= 2 \frac{d_{o}}{d_L} \sin \theta_{\rm JN}, \nonumber \\
\mathcal{A}_{2}^+ &= 3 \frac{d_{o}}{d_L} \sin^{2} \theta_{\rm JN}, \nonumber\\
\mathcal{A}_{2}^{\times} &= 0 \, . \label{eq:prec_amps}
\end{align}
In the approximation where the direction of total angular momentum is constant, the $\mathcal{A}^{+, \times}_{k}$
are time independent amplitudes, and the time dependence of the amplitude
functions is captured as a power series in the parameter $b = \tan(\beta/2)$.

Finally, we can use the harmonic decomposition in Eq.~(\ref{eq:cs_harms}) to obtain a decomposition
of the waveform, Eq.~(\ref{eq:prec_wf}),
\begin{eqnarray}
	h(t) &=& \mathrm{Re}
	\left[ \left( \frac{A_{o}(t) e^{2 i (\Phi_{S} + \alpha)}}{(1 + b^{2})^{2}} \right)
	\right. \nonumber \\
	&& \qquad \left.
	\sum_{k=0}^{4}  (b e^{-i\alpha})^{k}
	(F_{+} \mathcal{A}_{k}^{+} - i F_{\times} \mathcal{A}_{k}^{\times}) \right] \, .
	\label{eq:h_prec}
\end{eqnarray}
This allows us to clearly identify the impact of precession on the waveform.  First, precession
leads to an additional phase evolution at frequency $2 \Omega_{P}$ and a decrease in the amplitude
by a factor $(1 + b^{2})^{2}$.  The precessing waveform contains
five harmonics that form a power series in $b$, whose amplitude depends upon the
detector response, distance and viewing angle of the binary.  The frequency of each harmonic is offset from the
next by the precession frequency $\Omega_{P}$.  Similar results have been obtained previously, by manipulating
the spin-weighted spherical harmonic decomposition of the waveform, e.g.~\cite{o2020semianalytic, Lundgren:2013jla}.  
However, it was not previously observed that the relative amplitudes of the harmonics were related in a straightforward
manner.  In the Appendix, we present an alternative derivation of the result in Eq.~(\ref{eq:h_prec}) in terms of this 
spin-weighted spherical harmonic decomposition of the waveform, as is customary when producing waveform
models for precessing binaries \cite{Hannam:2013oca}.

As a final step, we would like to explicitly extract three more time-independent angles that characterize the
waveform, namely the polarization angle $\psi$, the initial phase $\phi_{o}$ and the initial polarization phase
$\alpha_{o}$.%
\footnote{The initial polarization phase $\alpha_{o}$ is sometimes denoted in the literature as $\phi_{JL}$.}

The unknown polarization $\psi$ is currently folded into the
detector response functions $F_{+, \times}$.  It is more useful to extract $\psi$ and then consider the
detector response to be a \textit{known} quantity dependent upon only the details of the detector and the direction to
the source.  Thus, we write the detector response as,
\begin{eqnarray}
	F_{+} &=& w_{+} \cos 2\psi + w_{\times} \sin 2\psi, \nonumber \\
	F_{\times} &=& - w_{+} \sin 2\psi + w_{\times} \cos 2\psi \label{eq:f_plus_cross},
\end{eqnarray}
where $w_{+}$ and $w_{\times}$ are the detector response functions in a fixed frame --- for a single detector it
is natural to choose $w_{\times} = 0$ and for a network to work in the dominant polarization for which
$w_{+}$ is maximized \cite{Harry:2010fr}.  The unknown polarization of the source relative to this preferred frame is denoted $\psi$.

To isolate the initial orbital and precession phases, we explicitly extract them from the binary's phase
evolution by introducing,
\begin{eqnarray}\label{eq:prec_phase}
	\Phi(t) &:=& \Phi_{S}(t) - \phi_{o} + \alpha(t) - \alpha_{o} \nonumber \\
	&=& \phi_{\mathrm{orb}}(t)- \phi_{o} + \int_{\alpha_{o}}^{\alpha(t)} \frac{2 b^{2}}{1 + b^{2}}  \, d \alpha \, .
\end{eqnarray}
Thus $\Phi(t)$ vanishes at $t=0$ and evolves as the sum of the orbital phase and an additional, precession
dependent, contribution.

We then substitute the expressions for $F_{+, \times}$, Eq.~(\ref{eq:f_plus_cross}), and $\Phi$, Eq.~(\ref{eq:prec_phase}),
into the expression for $h(t)$ given in Eq.~(\ref{eq:h_prec}), and
isolate the time-varying terms from the constant, orientation dependent angles. The waveform can be written
as the sum of five precessing harmonics, the amplitudes of which are constants that depend upon
the binary's sky location, distance and orientation:
\begin{equation}\label{eq:prec_harms}
	h = \sum_{k=0}^{4} w_{+} (h^{k}_{0} \mathcal{A}^{1}_{k} + h^{k}_{\frac{\pi}{2}} \mathcal{A}^{3}_{k})
		+ w_{\times} (h^{k}_{0} \mathcal{A}^{2}_{k} + h^{k}_{\frac{\pi}{2}} \mathcal{A}^{4}_{k}),
\end{equation}
where $h^{k}_{0, \frac{\pi}{2}}$ are the waveform harmonics and $\mathcal{A}_{k}^{\mu}$ are constants.
The waveform harmonics are
\begin{eqnarray}
	h^{k}_{0}(t) &=& \mathrm{Re} \left[ A_{o}(t) e^{2 i \Phi}
	\left( \frac{b^{k} e^{-i k(\alpha - \alpha_{o})} }{(1 + b^{2})^{2}} \right) \right],
			 \nonumber \\
	h^{k}_{\frac{\pi}{2}}(t) &=& \mathrm{Im} \left[ A_{o}(t) e^{2 i \Phi}
	\left( \frac{b^{k} e^{-i k(\alpha - \alpha_{o})}}{(1 + b^{2})^{2}} \right) \right] \, . \label{eq:td_harm}
\end{eqnarray}
The harmonics form a simple power series in $b e^{-i \alpha}$, so the amplitude of each successive
harmonic is reduced by a factor of $b$, and the frequency is reduced by $\Omega_{P}$.

The amplitudes for the harmonics are given by
\begin{align}
\mathcal{A}_{k}^1 &=\mathcal{A}_{k}^{+} \cos \phi_{k}\cos 2\psi
- \mathcal{A}_{k}^{\times}\sin \phi_{k}\sin 2\psi,
\nonumber \\
\mathcal{A}_{k}^2 &=\mathcal{A}_{k}^{+}\cos \phi_{k} \sin 2\psi
+ \mathcal{A}_{\times}\sin \phi_{k}\cos2\psi, \nonumber \\
\mathcal{A}_{k}^3 &= -\mathcal{A}_{k}^+\sin \phi_{k}\cos 2\psi
-\mathcal{A}_{k}^{\times} \cos \phi_{k}\sin 2\psi, \nonumber \\
\mathcal{A}_{k}^4 &=-\mathcal{A}_{k}^+\sin \phi_{k}\sin 2\psi
+\mathcal{A}_{k}^{\times}\cos \phi_{k}\cos 2\psi, \label{eq:prec_a_mu}
\end{align}
where the $\mathcal{A}^{+, \times}$ were introduced in Eq.~(\ref{eq:prec_amps}), $\psi$ is the polarization
and the phase angle for each harmonic is,
\begin{equation}\label{eq:harm_phase}
\phi_{k} = 2 \phi_{o} + (2 - k) \alpha_{o} \, .
\end{equation}
These amplitudes form a generalization of the $\mathcal{F}$-statistic decomposition of the non-precessing
binary waveform (see e.g. \cite{Harry:2010fr}).  In the limit that $b \rightarrow 0$, the
precessing decomposition reduces to the standard expression for the non-precessing waveform
as the amplitude for all harmonics other than $k=0$ vanish.

The precessing waveform can equally well be written in the frequency domain by performing a Fourier
transform of the time-domain expressions given above \cite{Droz:1999qx}.  In this case, Eq.~(\ref{eq:prec_harms})
is unchanged, as are the constant amplitude terms in Eq.~(\ref{eq:prec_a_mu}).
The frequency dependent harmonics are simply the
Fourier transform of the time-domain modes given in Eq.~(\ref{eq:td_harm}), and naturally satisfy
$h^{k}_{\frac{\pi}{2}} = i h^{k}_{0}$.

The expansion above is most natural when $b < 1$, which corresponds to opening angles of $\beta < 90^{\circ}$.
In cases where the opening angle is greater than $90^{\circ}$ it is natural to re-express the
waveform in terms of $c = b^{-1} = \cot(\beta/2)$
in which case the waveform can be expressed as a power series in $c$.
We will not discuss the large opening
angle calculation further in this
paper, but note that many of the arguments presented below would extend in a straightforward manner to
this case.

\subsection{Obtaining the harmonics}

Here, we give an explicit prescription to obtain the five harmonics for the waveform, introduced
in Eq.~(\ref{eq:prec_harms}).  To do so, we generate waveforms for orientations that contain only a subset of the harmonics,
and combine them to isolate a single harmonic.  For simplicity, we restrict attention to the $+$ polarization by
fixing $w_{+} = 1$, $w_{\times} = 0$ and consider a binary at a distance $d_{L} = d_{o}$.

\textit{Harmonics $k=0$ and $k=4$.}
When the viewing angle of the signal is aligned with the total angular momentum, $\theta_{\rm JN} = 0$, the observed
waveform contains only the zeroth and fourth harmonics as
$\mathcal{A}^{+, \times}_{1,2,3}$ vanish for $\theta_{\rm JN} = 0$.  We also fix $\alpha_{o} = 0$, to obtain,
\begin{eqnarray}
	h_{\phi_{o} = 0; \psi = 0}  &=& h^{0}_{0} + h^{4}_{0}, \nonumber \\
	h_{\phi_{o} = \tfrac{\pi}{4}; \psi = \tfrac{\pi}{4}}  &=& - h^{0}_{0} + h^{4}_{0}.
\end{eqnarray}
From these, we can extract the $k=0$ and 4 harmonics,
\begin{eqnarray}
	h^{0}_{0} &=& \tfrac{1}{2} \left( h_{\phi_{o} = 0, \psi = 0} - h_{\phi_{o} = \tfrac{\pi}{4}, \psi = \tfrac{\pi}{4} } \right), \nonumber \\
	h^{4}_{0} &=& \tfrac{1}{2} \left( h_{\phi_{o} = 0, \psi = 0} + h_{\phi_{o} = \tfrac{\pi}{4}, \psi = \tfrac{\pi}{4} } \right) \, .
\end{eqnarray}
The $\tfrac{\pi}{2}$ phases of the harmonics can be obtained in an identical way.

\textit{Harmonics $k=1$ and $k=3$.}
When the signal is edge on, the $\times$ polarization contains only the first and third harmonics.  Then, fixing
$\theta_{\rm JN} = \tfrac{\pi}{2}$ and $\psi = \tfrac{\pi}{4}$, we have,
\begin{eqnarray}
	h_{\alpha_{o} = 0; \phi_{o} = \tfrac{\pi}{4} } &=& - 2 \left( h^{1}_{0} + h^{3}_{0} \right), \nonumber \\
	h_{\alpha_{o} = \tfrac{\pi}{2}; \phi_{o} = 0 } &=& - 2 \left( h^{1}_{0} - h^{3}_{0} \right),
\end{eqnarray}
so that,
\begin{eqnarray}
	h^{1}_{0} &=& - \tfrac{1}{4} \left(h_{\alpha_{o} = 0; \phi_{o} = \tfrac{\pi}{4}} + h_{\alpha_{o} = \tfrac{\pi}{2}, \phi_{o} = 0} \right),
	 \nonumber \\
	h^{3}_{0} &=& -\tfrac{1}{4} \left(h_{\alpha_{o} = 0; \phi_{o} = \tfrac{\pi}{4}} - h_{\alpha_{o} = \tfrac{\pi}{2}, \phi_{o} = 0} \right).
\end{eqnarray}

\textit{Harmonic $k=2$.}
Finally, from the $+$ polarization of the edge-on waveform, we can extract the second harmonic --- in principle
we could also get $k=0$ and $k=4$, but we have already described a method to obtain them.
Fixing $\theta_{\rm JN} = \tfrac{\pi}{2}$ and $\psi = 0$ we have,
\begin{eqnarray}
	h_{\alpha_{o} = 0, \phi_{o} = 0} &=& \tfrac{1}{2} h^{0}_{0} + 3 h^{2}_{0} + \tfrac{1}{2} h^{4}_{0}, \nonumber \\
	h_{\alpha_{o} = \tfrac{\pi}{2}, \phi_{o} = 0} &=& - \tfrac{1}{2} h^{0}_{0} + 3 h^{2}_{0} - \tfrac{1}{2} h^{4}_{0},
\end{eqnarray}
so that,
\begin{eqnarray}
	h^{2}_{0} &=& \tfrac{1}{6} \left( h_{\alpha_{o} = 0, \phi_{o} = 0} +  h_{\alpha_{o} = \tfrac{\pi}{2}, \phi_{o} = 0 } \right).
\end{eqnarray}

\subsection{Precession with varying orientation}

\begin{figure*}[t]
	\begin{center}
	\includegraphics[width=0.49\textwidth]{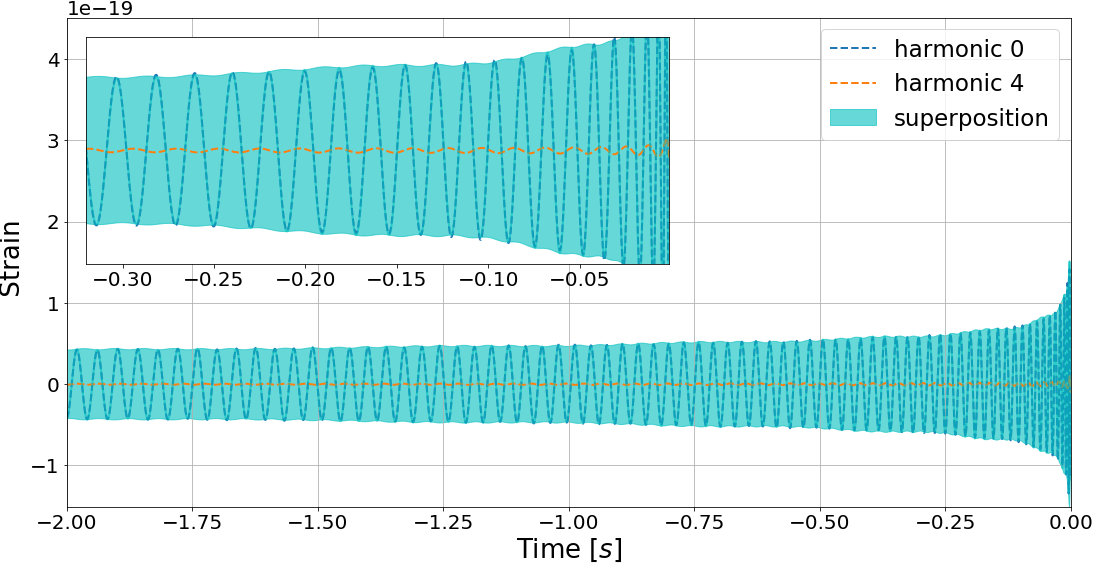}
	\includegraphics[width=0.49\textwidth]{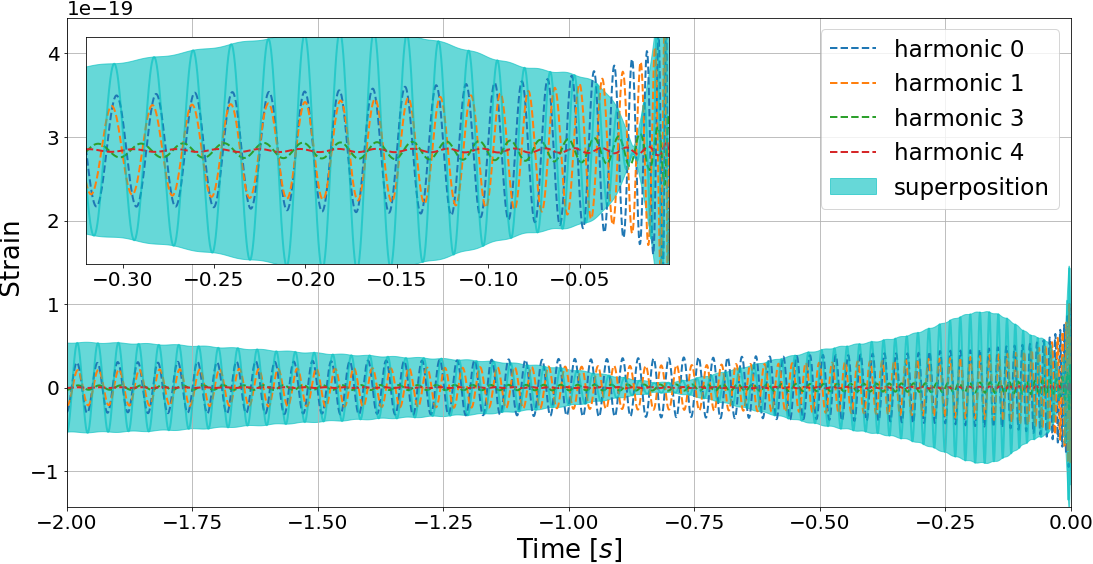}
	\includegraphics[width=0.49\textwidth]{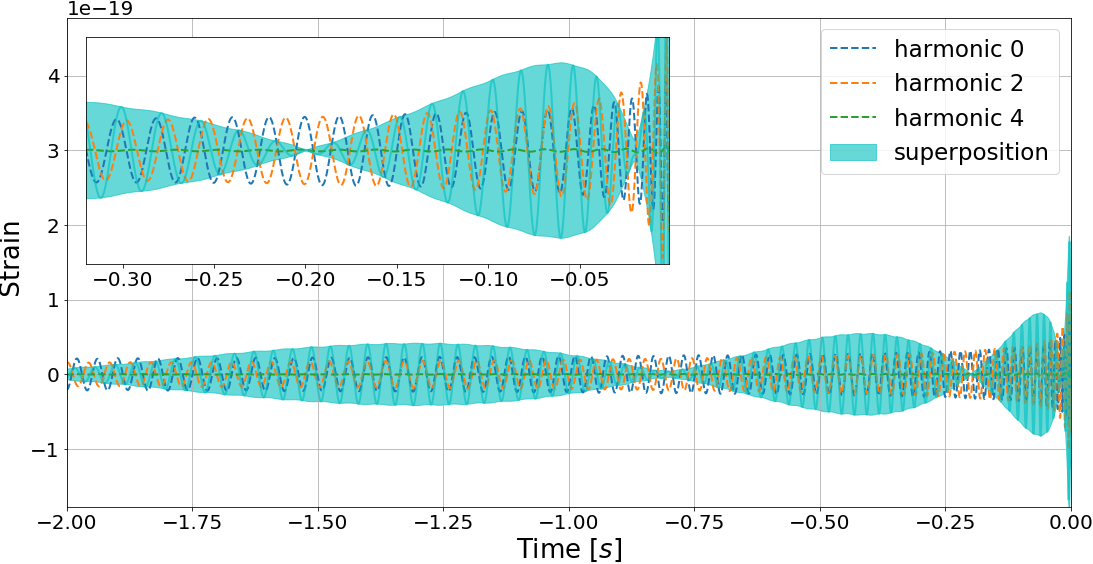}
	\includegraphics[width=0.49\textwidth]{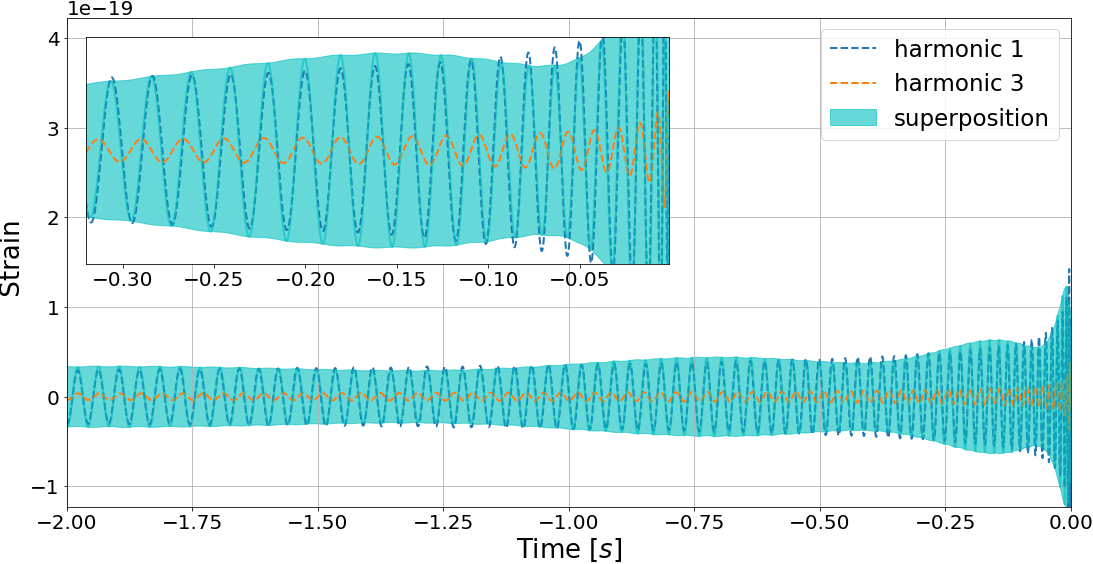}
	\end{center}
	\caption{The observed waveform from a $40 M_{\odot}$ binary with mass ratio $q = 6$,
	$\chi_{\mathrm{eff}} = 0$ and $\chi_{p} = 0.6$.
	The waveform is shown for four different binary orientations: $\theta_{JN}=0$ (upper left);
	$\theta_{JN}=45^{\circ}$, $\times$ polarization (upper right);
	$\theta_{JN}= 90^{\circ}$, $+$ polarization (lower left); $\theta_{JN}= 90^{\circ}$, $\times$ polarization (lower right).
	For each waveform, the harmonics that contribute
	to the signal, their sum and the envelope of the full precessing waveform are shown.  The insets show a zoom of a portion of the
	waveform to more clearly demonstrate that precession arises as a beating between the different harmonics.
	}
	\label{fig:harmonics}
\end{figure*}

The observable effect of precession will vary significantly with the binary orientation, as has been discussed in many previous
works, for example \cite{Apostolatos:1994mx, Brown:2012gs}.  Interestingly, both the amplitude and frequency of the observed 
precession depends upon the viewing angle.  The harmonic decomposition derived above provides a straightforward way to 
understand this effect.  The observed amplitude and phase modulations can be understood as the beating of the different 
harmonics against each other, with the amplitude of the composite waveform being maximum when the harmonics are in 
phase and minimum when they are out of phase. 

In Fig.~\ref{fig:harmonics}, we show the waveform for four different orientations: a) along $\mathbf{J}$, b) $\times$ polarization
at $45^{\circ}$ to $\mathbf{J}$, c/d) $+/\times$ polarization orthogonal to $\mathbf{J}$.   In all cases, we show the last two seconds 
of the waveform (from around 25 Hz) for a $40 M_{\odot}$ binary, with $q=6$, and in-plane spin on
the larger black hole of $\chi_P = 0.6$.  This configuration gives an opening angle of
 $\beta \approx 45^{\circ}$ (and $b\approx0.4$) which leads to significant precession effects in the waveform.
 
 When viewed along $\mathbf{J}$, there is minimal precession as only the $k=0$
and $4$ harmonics are present in the system and the $k=4$ harmonic is down-weighted by a factor of $b^{4} \approx 0.03$
relative to the leading harmonic.  Furthermore, the modulation comes from the beating of the 
$k=0$ and $k=4$ harmonics and occurs at four times the precession frequency.
When the line of sight is orthogonal to the total
angular momentum, the $k= 0, 2, 4$ harmonics are present in the $+$ polarized waveform and $k= 1,3$ in the
$\times$ polarization.  The $k=0$ and 2 harmonics have close to equal amplitude
(although $k=2$ is down-weighted by $b^{2} \approx 0.17$, the amplitude as given in Eq.~(\ref{eq:prec_amps}) is maximal).
Consequently the observed waveform has maximal amplitude and phase modulation due to precession.  For the $\times$ polarized 
signal, it is the $k=1, 3$ harmonics that contribute, with $k=3$ a factor of $b^{2} \approx 0.17$ smaller than $k=1$.  
Consequently, precession effects are less significant.  In both cases, precession occurs at twice the precession frequency 
as it is from the beating of  $k=0$ and $k=2$ ($+$ polarization) or $k=1$ and $k=3$
($\times$ polarization). For the $\times$ polarized signal with $\theta_{\rm JN} = 45^{\circ}$, the $k=0, 1, 3, 4$ harmonics are 
present, with  $k=0, 1$ dominating and having approximately
equal amplitude.  For this signal, the binary precesses from a face-on orientation, $\iota=0$
to edge-on, $\iota = 90^{\circ}$, and the waveform amplitude oscillates from the maximum to zero.  Here, modulations
occur at the precession frequency.

\subsection{Importance of precession over parameter space}

\begin{figure*}[t]
	\centering
	\includegraphics[width=0.99\textwidth]{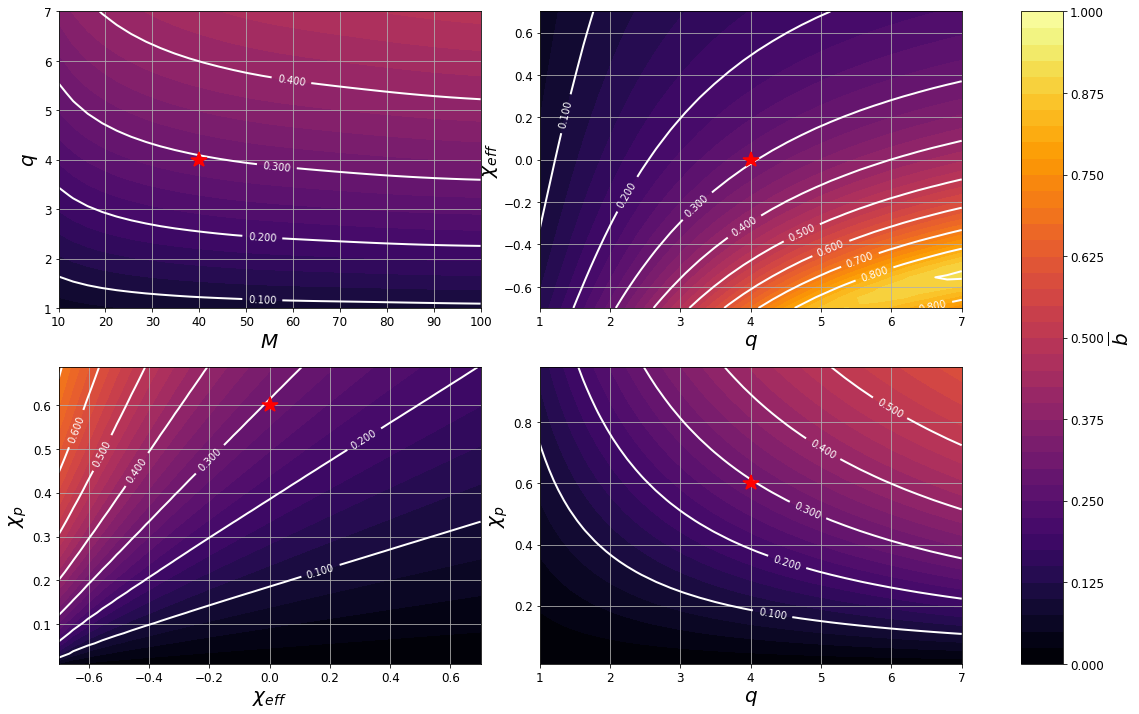}
	\caption{The value of $\overline{b}$ across the parameter space of total mass, mass ratio, $\chi_{\mathrm{eff}}$ and
	$\chi_{p}$.  In each figure, two of the parameters are varied while the other two are fixed to their fiducial values of
	$M = 40 M_{\odot}$, $q = 4$, $\chi_{\mathrm{eff}} = 0$, $\chi_{p} = 0.6$ (this point is marked with a $\star$ in all the plots).
	The total mass has a limited impact on the value of $\overline{b}$, for masses over $M \approx 40 M_{\odot}$; below this
	the $\overline{b}$ increases with mass, as the later parts of the merger are brought into the most sensitive band of the
	detector.
	The value of $\overline{b}$ is seen to increase as the mass ratio or precessing spin $\chi_{p}$ are increased
	and decrease as the aligned component of the spin $\chi_{\mathrm{eff}}$ increases.  
	Thus, the value of $b$ is largest for a binary with unequal masses, a large spin on the
	more massive component which has significant components both in the plane of the orbit and anti-aligned with the
	orbital angular momentum.}
	\label{fig:variation_of_b}
\end{figure*}

From the intuitive discussion of precession presented in \cite{Apostolatos:1994mx, Buonanno:2002fy, Brown:2012gs} and
summarized in Section \ref{sec:precession}, it is straightforward to identify regions of parameter space where precession is most
likely to have a significant impact upon the binary dynamics and, consequently, the observed waveform.  Specifically,
we expect that higher mass ratios, larger in-plane spins and negative aligned spin components will all lead to a larger
opening angle and more significant precession \cite{Brown:2012gs}.  Here we briefly revisit this discussion, framing our results in terms
of the parameter $b$ introduced earlier.  Explicitly, we introduce the waveform-averaged value of $b$ as,
\begin{equation}\label{eq:b_bar}
\overline{b} := \frac{ |h^{1} | }{ | h^{0} |} = \sqrt{\frac{\int df \frac{ |h_{1}|^{2}}{S_{n}(f)}}{\int df \frac{ |h_{0}|^{2}}{S_{n}(f)}}} \, ,
\end{equation}
where $h^{0,1}$ are the harmonics of the waveform introduced in Eq.~(\ref{eq:td_harm}) and $S_n(f)$ is the noise power 
spectrum of the detector.  For this work, we choose $S_n(f)$ to be the design-sensitivity Advanced 
LIGO noise curve~\cite{LIGOScientific:2018mvr} and evaluate the integral over the frequency range 
$f \in [20,1024]$\,Hz
\footnote{Using a realistic noise curve similar to the observed curves during 01 and O2 would change the reported values slightly,
as these noise curves are less sensitive than design, particularly at low frequencies. The qualitative patterns seen in the figure would remain the same however}.   For binaries where the opening angle $\beta$ is approximately constant, $\overline{b} \approx \tan(\overline{\beta}/2)$.

Fig.~\ref{fig:variation_of_b} shows the value of $\overline{b}$ on several two-dimensional slices through the
four dimensional parameter space of total mass $M$,  mass ratio $q$, effective spin $\chi_{\mathrm{eff}}$ and
precessing spin $\chi_{p}$.  Keeping other quantities fixed, the value of $\overline{b}$ increases with total mass.
For higher masses, the late inspiral and merger occur in the sensitive band of the detectors and, close to merger, the opening
angle increases as orbital angular momentum is radiated.
For masses above $40 M_{\odot}$ the mass dependence of $\overline{b}$ is small, with only a $10\%$
decrease from $40 M_{\odot}$ to $100 M_{\odot}$.  Thus, for the other figures,
we fix $M = 40 M_{\odot}$ and investigate the dependence of $\overline{b}$ on $q, \chi_{\mathrm{eff}}$
and $\chi_{p}$.
The dependence of $\overline{b}$ follows directly from Eq.~(\ref{eq:opening_angle}).  The opening angle will increase
with  mass ratio, as the orbital angular momentum decreases.  The opening angle, and also $\overline{b}$, increase
with $\chi_p$.  It follows directly from the definition that $\tan{\beta}$ scales linearly with $\chi_{p}$, and hence approximately
linearly for $b = \tan (\beta/2)$.  Finally, the opening angle decreases as
the effective spin $\chi_{\mathrm{eff}}$ increases, so that the largest value of $\overline{b}$ is obtained with
significant spin anti-aligned with $\mathbf{J}$.

Over much of the parameter space we have explored, $\overline{b} \lesssim 0.3$.  This includes
binaries with mass ratio up to 4:1, with precessing spin $\chi_{p} \lesssim 0.6$, and zero or positive aligned spin,
$\chi_{\mathrm{eff}} \ge 0$.  Only a small part of parameter space
has $\overline{b} > 0.4$, the value used in generating the waveforms in Figure \ref{fig:harmonics},
and $b > 0.5$ is only achieved with at least
two of: a) close to maximal $\chi_{p}$, b) high mass ratio, $q \gtrsim 5$ or c) significant spin anti-aligned with the orbital angular momentum $\chi_{\mathrm{eff}} \lesssim -0.4$.

\begin{figure}
	\centering
	\includegraphics[width=0.49\textwidth]{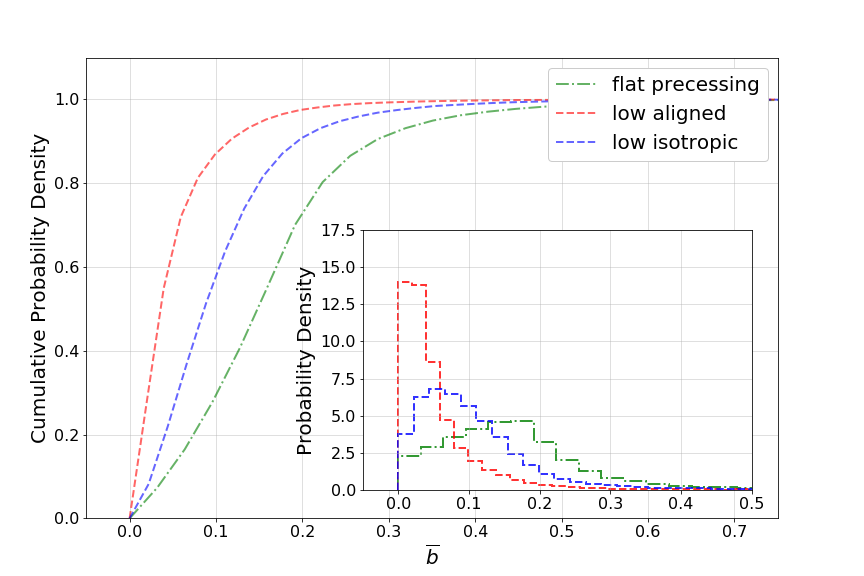}
	\caption{The distribution of $\overline{b}$ for a 3 different populations of binary black holes. Each
	population assumes either a low-isotropic, low-aligned or a flat precessing spin distribution. A 
	power-law distribution in masses is assumed in all cases (see text for details).}
	\label{fig:b_population}
\end{figure}

Next, we consider the importance of precession for an astrophysically motivated population.  
In Fig.~\ref{fig:b_population}, we show the distribution of $\overline{b}$ for three distributions of black hole 
masses and spins. For each population, we generate 100,000 binaries uniformly in co-moving distance, 
with masses drawn from a power law distribution ---
$p(m_{1}) \propto m_{1}^{-\alpha}$, with $\alpha = 2.35$ --- and different spin distributions, which are the same as those used in 
Refs.~\cite{Farr:2017uvj, Tiwari:2018qch, fairhurst2019will}. We consider populations where the spins are preferentially 
low and aligned with the binary orbit; low and isotropically aligned or drawn from a flat distribution and preferentially leading to 
precession.  A low spin distribution is a triangular distribution peaked 
at zero spin and dropping to zero at maximal spin while a flat distribution is a uniform between zero and maximal spin.
The \emph{aligned} distribution is strongly peaked towards aligned spins, while the \emph{isotropic} distribution assumes a uniform distribution of spin orientations over the sphere.
The \emph{precessing} distribution is strongly peaked towards spins orthogonal to the orbital angular 
momentum, i.e., with significant orbital precession~\cite{Antonini:2017tgo, Rodriguez:2018jqu}. 
To account for observational biases, we keep only those signals that would be observable above a fixed threshold in a 
gravitational wave detector. We find that even for the most extreme precessing 
population considered, the mean value of $\overline{b}$ is 0.15 with over 90\% of binaries having $\overline{b} < 0.3$. 
This result is obviously sensitive to the assumptions on the mass and spin distribution.  
In Ref.~\cite{fairhurst2019will} we investigate a larger number of spin distributions, including ones which allow 
for large spin magnitudes, and we find that the peak of the $\overline{b}$ distribution is below $0.2$ and that
over 90\% of binaries  have $\overline{b}  < 0.4$ in all cases.

\begin{figure*}[t]
	\centering
	\includegraphics[width=0.99\textwidth]{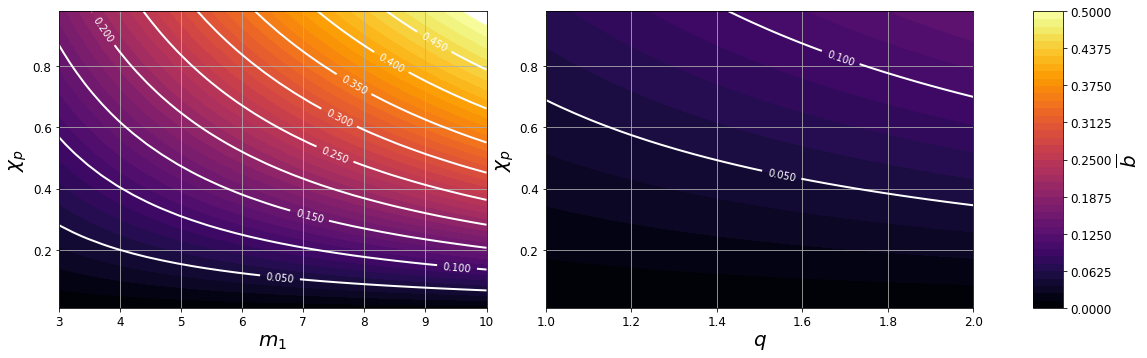}
	\caption{
	The value of $\overline{b}$ across the binary neutron star and neutron-star--black-hole space.  The left figure
	shows the variation of $\overline{b}$ for an NSBH system with a $1.4 M_{\odot}$ neutron star, $\chi_{\mathrm{eff}} = 0$
	and varying black hole mass and $\chi_{p}$.
	The right figure shows the variation of $\overline{b}$ against mass ratio and $\chi_{p}$ for a binary neutron star system
	of total mass $2.7 M_{\odot}$ and $\chi_{\mathrm{eff}} = 0$.
	}
	\label{fig:bns_nsbh}
\end{figure*}

Fig.~\ref{fig:bns_nsbh} shows $\overline{b}$ for a range of neutron star--neutron star and neutron
star--black hole binaries.  For neutron star--black hole binaries, the picture is similar to that for black hole binaries, with
large values of $\overline{b}$ observed for high mass ratios and large $\chi_{P}$.  However, as an earlier part of the
waveform is in the detector's sensitive band, the impact of precession is less observable at fixed mass ratio than for
higher mass black hole binaries.  For neutron star binaries, the value of $\overline{b}$ remains below $0.15$ across the
parameter space, and is less than $0.05$ for reasonable neutron star spins, $\chi \lesssim 0.4$.

\section{The two-harmonic approximation}
\label{sec:twoharmonics}

The precessing waveform can be expressed as the sum of five harmonics whose amplitudes form
a power series in $b = \tan(\beta/2)$.
Furthermore, over the majority of the space of binary mergers, the value of $b$ is less than $0.3$.
In addition, for $\overline{b} \le 0.4$ the dominant harmonic --- the one containing the most power ---
must be either $k=0$ or $1$.
Thus, for the vast majority of binary mergers, we expect that these two harmonics will be the most significant.

This motivates us to introduce the \textit{two-harmonic approximation}, in which we generate a waveform
containing only the $k=0$ and $k=1$ harmonics, i.e.,
\begin{equation}
	h =
	\sum_{k=0, 1} w_{+} (h^{k}_{0} \mathcal{A}^{1}_{k} + h^{k}_{\frac{\pi}{2}} \mathcal{A}^{3}_{k})
		+ w_{\times} (h^{k}_{0} \mathcal{A}^{2}_{k} + h^{k}_{\frac{\pi}{2}} \mathcal{A}^{4}_{k}) \, .
\end{equation}

The expression for the two-harmonic waveform can be simplified by restricting to the single detector
case (i.e., setting $w_{+} = 1$ and $w_{\times} = 0$), explicitly working with the waveform in the frequency domain,
for which $h^{k}_{\tfrac{\pi}{2}}(f) = i h^{k}_{0}(f)$, and dropping the subscript $0$ on the zero-phase
waveform, so that $h^{k}(f) := h^{k}_{0}(f)$.  The two harmonics of interest are,
\begin{align}
	h^{0}(f) &=A_{o}(f) e^{2 i \Phi(f)} \left( \frac{1}{(1 + b(f)^{2})^{2}} \right), \\
	h^{1}(f) &=A_{o}(f) e^{2 i \Phi(f)} \left( \frac{b(f) e^{-i (\alpha (f)- \alpha_{o})} }{(1 + b(f)^{2})^{2}} \right),
\end{align}
and the two-harmonic waveform then becomes,
\begin{equation}\label{eq:2harm}
	h_{\mathrm{2harm}} = \mathcal{A}_{0} h^{0} + \mathcal{A}_{1} h^{1} \, ,
\end{equation}
where,
\begin{align}\label{eq:2harm_amps}
	\mathcal{A}_{0} =& \frac{d_0}{d_L} \left(\frac{1+\cos^2 \theta_{\rm JN}}{2} \cos 2\psi - i \cos \theta_{\rm JN} \sin 2\psi \right) \times \nonumber \\
	& e^{-i  (2 \phi_{o} + 2 \alpha_{o} )}, \nonumber \\
	\mathcal{A}_{1} =&\frac{d_0}{d_L}  \left(\sin 2\theta_{\rm JN}\cos 2\psi - 2 i \sin \theta_{\rm JN} \sin 2\psi \right) \times \nonumber \\
	& e^{-i (2 \phi_{o} + \alpha_{o} )}.
\end{align}
Thus, the two-harmonic waveform is composed of two components that have frequencies
offset by $\Omega_{P}$, and any observed amplitude and phase modulation of the
waveform is caused by the beating of one waveform against the other.  The relative
amplitude and phase of the two harmonics is encoded by %
\begin{eqnarray}\label{eq:zeta}
	\zeta & := & \frac{\overline{b} \mathcal{A}_{1}}{\mathcal{A}_{0}} \\
	& = & \overline{b} e^{i \alpha_{o}}  \, \left(\frac{
	\sin2\theta_{\rm JN} \cos 2\psi - 2 i \sin \theta_{\rm JN} \sin 2\psi  }{
	\tfrac{1}{2}(1 +\cos^2 \theta_{\rm JN}) \cos 2\psi - i \cos \theta_{\rm JN} \sin 2\psi }\right)  \, . \nonumber
\end{eqnarray}
The value of $\zeta$ depends upon the viewing angle, encoded
in $\theta_{\rm JN}$ and $\psi$, and the initial precession phase $\alpha_{o}$.
It is not difficult to show that $\zeta$ can take any value as the parameters $\theta_{\rm JN}$, $\psi$, $\alpha_{o}$
are varied.  For example, at $\theta_{\rm JN} = 0$,
$\mathcal{A}_{1}$ vanishes and so does $\zeta$, while at $\theta_{\rm JN} = \pi/2$ and $\psi = \pi/4$, $\mathcal{A}_{0}$
vanishes and $\zeta \rightarrow \infty$.  Since the initial precession phase $\alpha_{o}$ is a free parameter, the
phase of $\zeta$ also can take any value.
The overall amplitude and phase of the signal also depends upon the distance and coalescence phase so that
any values of the amplitude and phase of the signal in the two harmonics are consistent with a signal.

\section{Validity of the two-harmonic waveform}
\label{sec:validity}

To investigate the validity of the two-harmonic approximation, we compare the approximate waveform with
the full, five-harmonic, precessing waveform across the parameter space.  The error will be of order $b^{2}$, which is
small over much of the parameter space, and for the majority of orientations.

\begin{figure*}[t]
	\begin{center}
	\includegraphics[width=0.99\textwidth]{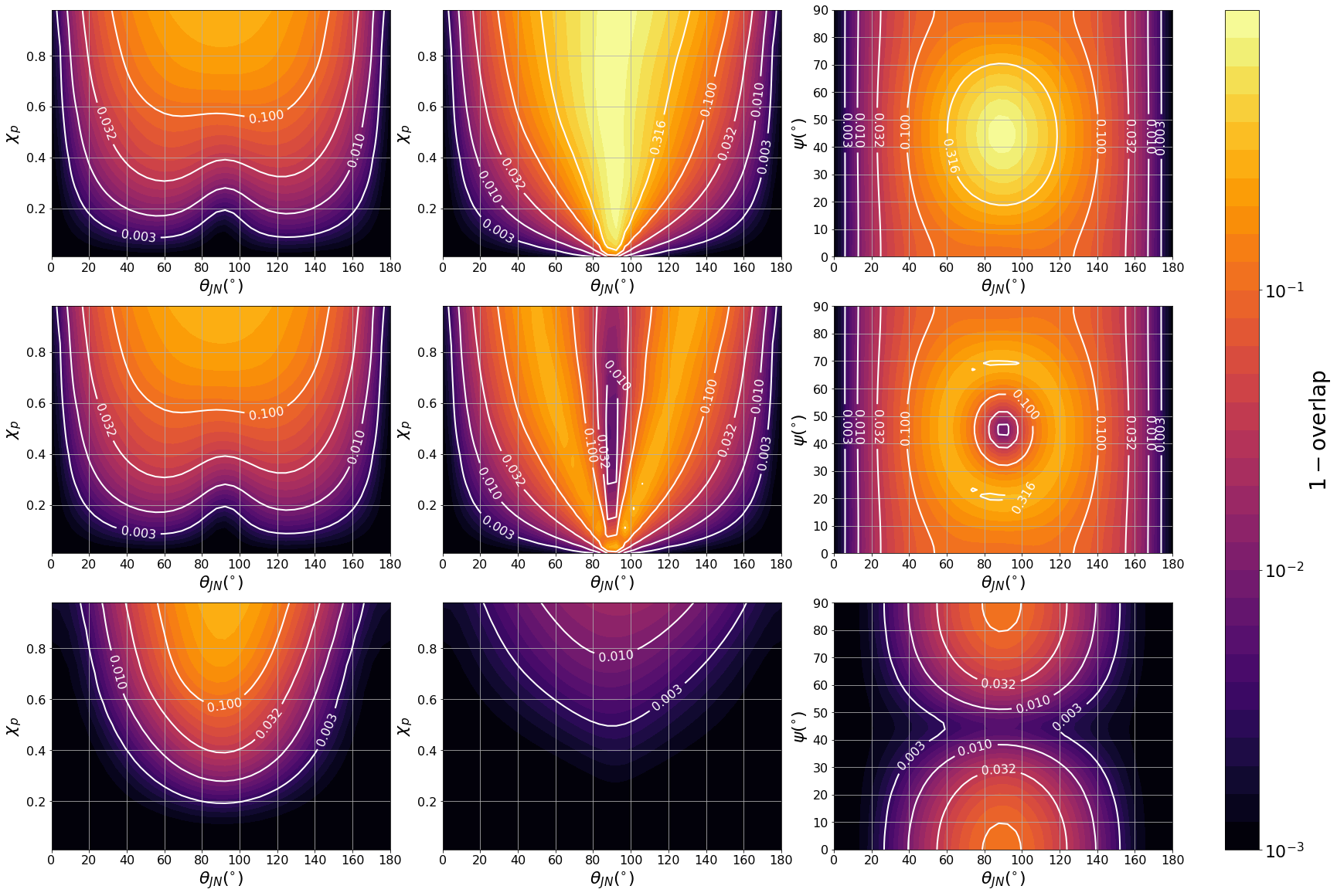}
	\end{center}
	\caption{The overlap between a precessing waveform and a subset of the harmonics, as a function of the
	precessing spin and binary orientation for a $40 M_{\odot}$ binary with mass ratio $q=4$ and $\chi_{\mathrm{eff}}= 0$.
	The top row shows the overlap between the leading, $k=0$, harmonic and
	the full waveform; the second row shows the overlap between the dominant harmonic and the
	full waveform; the bottom row shows the overlap between our two-harmonic precessing waveform and the full waveform.
	The first column is for the $+$ polarization, second for $\times$ and third for fixed $\chi_{P} = 0.6$ and varying
	polarization.
	}
	\label{fig:overlap_chip_theta}
\end{figure*}

Fig.~\ref{fig:overlap_chip_theta} shows the overlap between the full waveform and a subset of the harmonics for a binary
with $M = 40 M_{\odot}$, $q=4$ and $\chi_{\mathrm{eff}}= 0$, while varying the orientation and value of $\chi_{P}$.  In each
case, we calculate,
\begin{equation}
O(h, h')= \frac{\mathrm{max}_{\phi_{o}} (h | h')} {|h| | h' |},
\end{equation}
where,
\begin{equation}
	(a | b) = 4 \, \mathrm{Re} \int_{f_{o}}^{\infty} \frac{a^{\star}(f) b(f)}{S(f) }df,
\end{equation}
and $S(f)$ is the power spectral density of the detector data. 
Thus the overlap is maximized over the phase, but not over time or any of the mass and
spin parameters.  An overlap of close to unity shows that the two waveforms are very similar, while a lower value of
overlap implies significant deviations between the waveforms.  As a rule of thumb, an overlap $O(h, h') \lesssim 1 - 3/\rho^{2}$
will be observable at a signal to noise ratio $\rho$ \cite{Flanagan:1997kp, Miller:2005qu, Baird:2012cu}.

We calculate the overlap of the full waveform, $h$, against
\begin{enumerate}
\item the leading order waveform in the precession expansion, $h^{0}$;
\item the dominant harmonic, i.e. the harmonic of $h^{0}$ and $h^{1}$ which contains the largest fraction of the power
in the full waveform;
\item the two-harmonic waveform with the appropriate values of $\mathcal{A}_{0}$ and $\mathcal{A}_{1}$.
\end{enumerate}

For the $+$ polarized waveform (left column), the $k=0$ harmonic is dominant for all values of $\theta_{\rm JN}$ and $\chi_{P}$, so
that the observed overlap with the full waveform is above $0.8$ across the parameter space.  For $\theta_{\rm JN} \approx 0$ 
or small values of $\chi_{P}$,
the other harmonics make a minimal contribution and the overlap is close to unity.  For larger
values of $\theta_{\rm JN}$ and $\chi_{P}$ the other harmonics are more significant and the overlap drops to $0.9$ or less.
The two-harmonic waveform is a significantly better match to the full waveform, with an overlap greater than $0.99$ for
much of the parameter space, and only below $0.9$ for edge-on systems with high $\chi_{P}$ where the $k=2$
harmonic contributes most strongly (and the $k=1$ contribution vanishes).

For the $\times$ polarized waveform (center column), the effect of incorporating the $k=1$ harmonic is dramatic.
For $\theta_{\rm JN} = 90^{\circ}$ the $k=0$ contribution vanishes and only the $k=1, 3$ harmonics are present.  Thus,
the overlap with harmonic $k=0$ is essentially zero.  Using the best of $k=0, 1$ provides a good overlap with the
edge-on waveform, but there is still a poor overlap at $\theta_{\rm JN} \approx 60^{\circ}$ where both the $k=0$ and $1$
harmonics contribute significantly to the waveform. This effect has been observed previously, for example in \cite{Brown:2012gs,
o2020semianalytic} and a geometric understanding of its origin provided.  
The two-harmonic waveform matches remarkably well to the full waveform,
with the largest differences for $\theta_{\rm JN} = 90^{\circ}$ and $\chi_{P} \approx 1$ where the overlap drops to $0.99$
due to the contribution from the $k=3$ harmonic.

The right column shows the overlap as the orientation of the binary changes.  As expected, at points where
the $k=0$ harmonic vanishes ($\theta_{\rm JN} = 90^{\circ}$ and $\psi = 45^{\circ}$), the overlap with this
harmonic drops to zero.  The dominant harmonic is a good match to the waveform, except for orientations where
two harmonics contribute significantly.  As discussed in detail in Ref.~\cite{Brown:2012gs}, this 
corresponds to configurations where the binary orientation passes through the null of the detector response
(i.e.~the signal goes to zero) once per precession cycle.
Thus, the radius of the circle with poor overlaps is approximately equal to the opening angle of the binary.  The two-harmonic
approximation provides an excellent fit to the full waveform over the majority of orientations, only
 dropping below $0.95$ for orientations where $\theta_{\rm JN} \rightarrow 90^{\circ}$ and $\psi \approx 0, 90^{\circ}$, where
 the $k=2$ harmonic is most significant.

\begin{figure}[t]
	\begin{center}
	\includegraphics[width=0.49\textwidth]{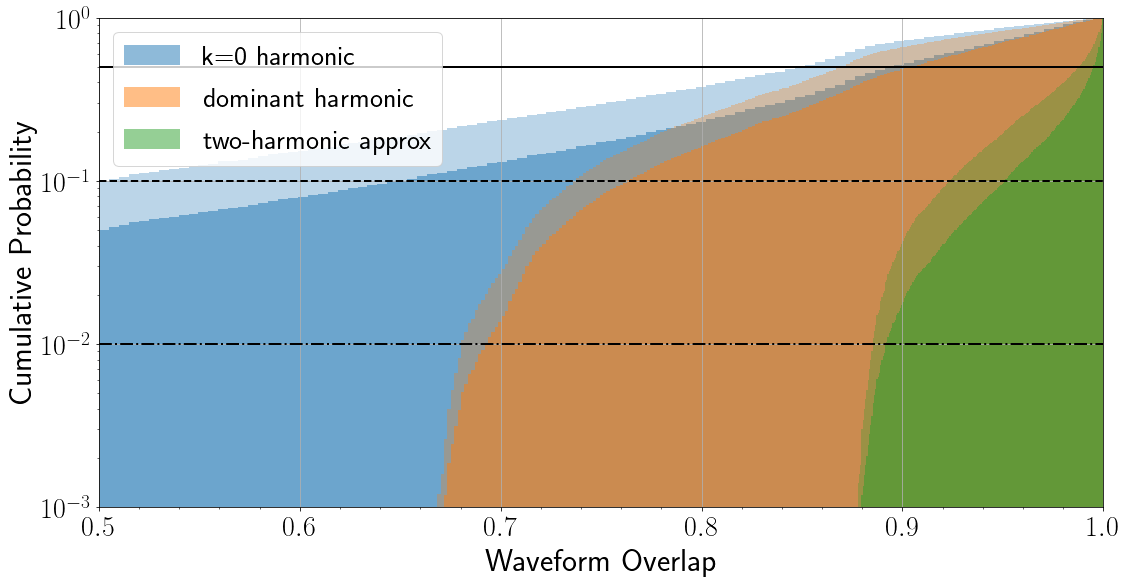}
	\includegraphics[width=0.49\textwidth]{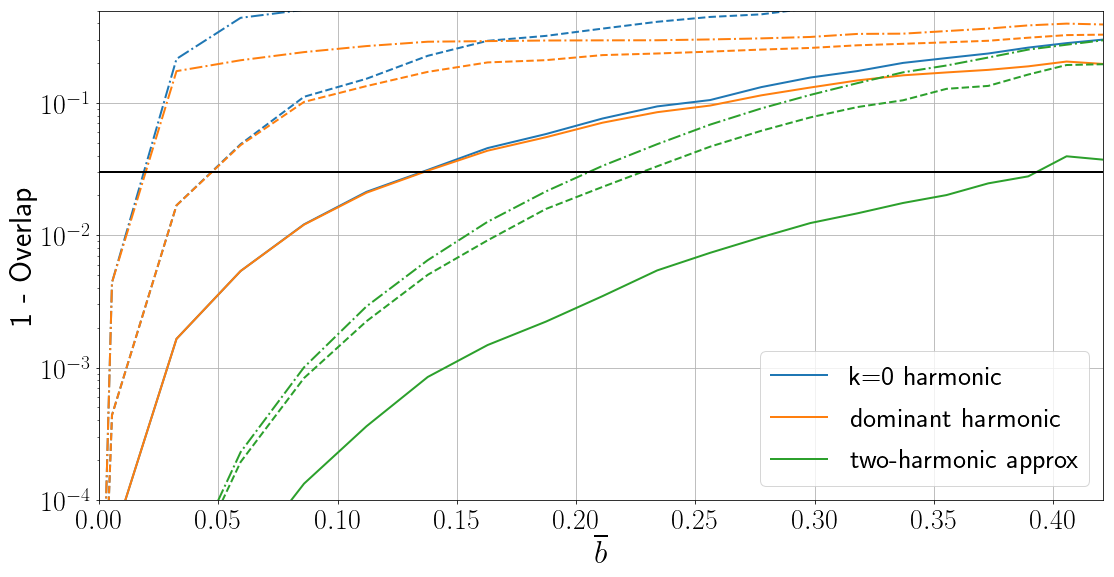}
	\end{center}
	\caption{
	The distribution of the overlap of the precessing waveform with the $k=0$, dominant and two-harmonic waveforms for
	a population of signals with $M = 40 M_{\odot}$, $q = 4$, $\chi_{\mathrm{eff}} = 0$.  The top plot shows
	the overlap distribution for $\chi_{P} = 0.6$, with random orientation of the signal.
	The lighter shaded regions give the distribution for a randomly oriented population of sources and the darker
	regions for the expected observed distribution (for a uniform-in-volume source).  The lower plot shows the
	overlap between full and approximate waveforms as a function of $\overline{b}$.  The lines
	on the plot show the value of the overlap for the median (solid line), worst 10\% (dashed)
	and worst 1\% (dot-dashed) of signals.
	}
	\label{fig:pop_overlap}
\end{figure}

Next, we investigate the validity of the two-harmonic approximation for a population of binaries.
To begin with, let us fix the masses and spins and just consider
the effect of binary orientation.  As before, we choose $M = 40 M_{\odot}$, $q = 4$, $\chi_{\mathrm{eff}} = 0$ and $\chi_{P} = 0.6$,
corresponding to $\overline{b} \approx 0.3$,
with the binary orientation distributed uniformly over $\cos(\theta_{\rm JN}), \phi_{o},
\alpha_{o}, \psi$.  Fig.~\ref{fig:pop_overlap} shows the distribution of the overlap between the full waveform and
1) the $k=0$ harmonic, 2) the dominant harmonic and 3) the two-harmonic approximation.  The results are shown for both a 
uniformly distributed population, and a population of signals observable above a fixed threshold in the detector --- thereby favoring 
orientations that produce the largest amplitude gravitational
wave.  The median overlap with either the $k=0$ or dominant harmonic is $\lesssim 0.9$, while the two-harmonic approximation 
improves the median overlap to $0.99$.  Using the dominant harmonic, there are a small fraction of signals with overlaps
of $0.7$ or lower (and for the $k=0$ harmonic, this tail extends to overlaps of 0.2), while for the two harmonic approximation, the
worst overlap is $0.88$.

We can use these results to obtain a \textit{rough} sense of the benefits of performing a search using the two-harmonic
approximation.  Previous, more detailed, investigations of this question have been carried out in, 
e.g.~\cite{Harry:2016ijz, Brown:2012gs, Ajith:2012mn}. Current gravitational wave searches make use of spin-aligned waveforms 
\cite{Messick:2016aqy, Usman:2015kfa}, and a precessing waveform
will naturally be identified by a spin-aligned waveform which matches well the dominant harmonic.  
Thus, we can use the
overlaps between the precessing waveforms and dominant harmonics as a proxy for the performance of an aligned spin
search.  Since the median overlap is $0.9$ we would expect to recover approximately 70\% as many signals ($\approx 0.9^{3}$ 
for a population uniform in volume) as with a full precessing search, above a fixed threshold.  A search based upon the 
two-harmonic approximation would recover around 97\% of these signals, indicating an improvement of over 30\%
in sensitivity to such systems.

We also show how the distribution of overlaps varies across the mass and spin parameter space, as encoded by the
parameter $\overline{b}$ and plotted for three choices of spin distribution in Figure \ref{fig:b_population}.%
\footnote{While these
plots were made with fixed masses and $\chi_{\mathrm{eff}}$, they should give a reasonable indication of the accuracy 
of the two-harmonic waveform across the mass and spin parameter space, as a function of $\overline{b}$.  
For different masses and spins, the evolution of the precession angle during the coalescence can have a slight 
impact upon the relative importance of the modes but, as $b$ typically does not change significantly over the observable 
waveform, this effect is likely to be small.
Furthermore, as different modes are not perfectly orthogonal, the degree to which they are not will also have a small
effect upon the results.  As shown in Section \ref{sec:obs_prec}, the harmonics are close to orthogonal
for $M \lesssim 40 M_{\odot}$ so that the results shown here will be representative, at least at lower masses.}
For $\overline{b}
\lesssim 0.13$ --- accounting for three quarters of signals in the low-isotropic population ---  the median overlap between the 
dominant harmonic and the full waveform is above $0.97$.  Thus, for the majority of expected signals, the spin-aligned
search will have good sensitivity.  However, even for low values of $\bar{b}$ there will be some orientations of signals
where two dominant harmonic will not match the waveform well, while the two-harmonic waveform still provides an 
essentially perfect representation of the waveform for all orientations.
At $\overline{b} \approx 0.25$ the median overlap with the  dominant harmonic waveform drops to $0.9$, and it
is here that a search with the two-harmonic approximation could provide a 30\% improvement.  We note, however,
that for the low-isotropic distribution this accounts for only 5\% of systems.
While systems 
with such significant precession may be rare they would come from interesting areas of parameter space, with
high mass ratios and spins. 
It is only at $\overline{b} = 0.4$ that the median overlap for the two harmonic waveform drops to 0.97, indicating
a 10\% loss relative to an ideal search, but also 70\% improvement over a spin-aligned search.  

\section{Searching for precessing binaries}
\label{sec:prec_search}

The two-harmonic approximation provides an ideal basis to develop a search for binaries
with precession.  The typical approach to searching for binary coalescences has been to generate a template-bank
of waveforms that covers the parameter space \cite{Owen:1995tm, Owen:1998dk, Babak:2006ty}.
These templates comprise discrete points in the
mass and spin space chosen so that the waveform produced by a binary anywhere in the parameter
space of interest has a match of at least 97\% with one of the templates.
The waveform for each template is then match-filtered against the data to identify peaks of high SNR, and
various signal consistency and coincidence tests are used to differentiate signals from non-stationary noise
transients \cite{Allen:2005fk, Allen:2004gu, Babak:2012zx, Messick:2016aqy, Usman:2015kfa}.  Current searches
make use of a template bank covering the four dimensional mass and aligned-spin space \cite{DalCanton:2017ala,
Mukherjee:2018yra}.%
\footnote{As we have discussed, the most significant effect on the observed waveform arises due to the
effective spin $\chi_{\mathrm{eff}}$, which is a combination of the aligned spin components of the two waveforms.
Thus, although the template space is four dimensional, one of the spin directions provides limited variation to the
waveforms, and thus is relatively straightforward to cover.}
The search takes advantage of the fact that changing the sky location, distance and orientation of the binary
only changes the overall amplitude and phase of the signal, and these quantities can be maximized over in
a simple manner.

When developing a search for precessing binaries, the search becomes more challenging due to the
increasing number of parameters.  In principle, it is necessary to search over two masses and
six spin components, although, in practice it will probably be sufficient to restrict to the masses, $\chi_{\rm eff}$
and $\chi_{P}$.  The second complication is that the observed morphology of the waveform varies as
the orientation of the binary changes, and it becomes necessary to search over binary orientation $\theta_{\rm JN}$,
polarization $\psi$ and precession phase $\alpha_{o}$, although methods have been developed to
straightforwardly handle a subset of these parameters \cite{Pan:2003qt, Harry:2016ijz}.

The two-harmonic waveform can be used to maximize the SNR over the binary orientation in a simple way.
The two complex amplitudes $\mathcal{A}_{0}$ and $\mathcal{A}_{1}$, defined in Eq.~(\ref{eq:2harm_amps}),
are dependent upon five variables:
the distance, $d_{L}$, binary orientation, $\theta_{\rm JN}$, $\psi$, and the initial orbital and precession phases,
$\phi_{o}$, $\alpha_{o}$.  Since $\mathcal{A}_{0}$ and $\mathcal{A}_{1}$ can take any value in
the complex plane, it is possible to construct the two-harmonic SNR by filtering the two harmonics $h_{0}$
and $h_{1}$ against the data and then freely maximizing the amplitudes so that,
\begin{equation}\label{eq:rho_2harm}
	\rho_{\mathrm{2harm}}^{2} = \rho_{0}^{2} + \rho_{1}^{2} \, .
\end{equation}
If the harmonics are not orthogonal, the two-harmonic SNR should be calculated using $h^{0}$ and
$h_{\perp}^{1}$ --- the $k=1$ harmonic with any component proportional to $h^{0}$ removed.
The extrinsic
parameters of the binary (distance, sky location, orientation, orbital and precession phase)
can be searched over through maximization over the amplitudes of the two harmonics, leaving only
the masses and spins as dimensions to search using a bank of waveforms.

\begin{figure*}[t]
	\begin{center}
	\includegraphics[width=0.99\textwidth]{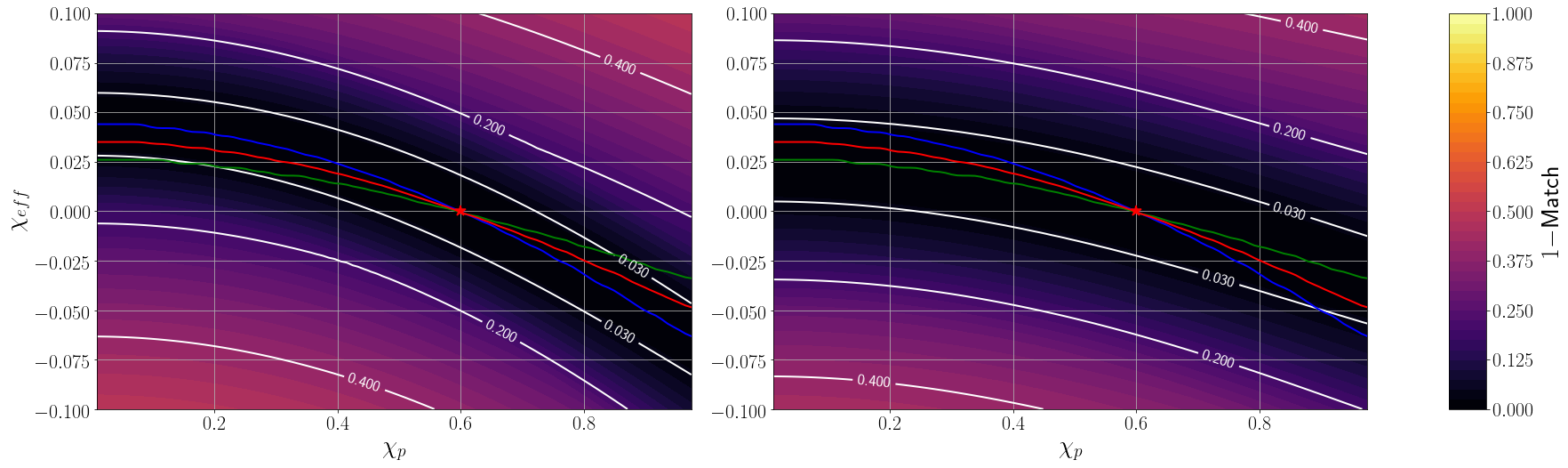}
	\end{center}
	\caption{The mismatch between the $k=0$ (left) and $k=1$ (right) harmonic of two precessing signals as the effective
	spin $\chi_{\mathrm{eff}}$ and precessing spin $\chi_{P}$ are varied.  For all waveforms, the total mass is fixed to
	$40 M_{\odot}$ and the mass ratio to 4.  One waveform has $\chi_{\mathrm{eff}} = 0$ and $\chi_{P} = 0.6$ (the
	point marked by a star), while the spins of the second waveform are varied.  The blue and green lines show the
	value of $\chi_{\mathrm{eff}}$, for the $k=0$ and $k=1$ harmonics respectively, which gives the largest match with the 
	fiducial waveform; the red line is the average of these values.
	}
	\label{fig:chi_eff_chi_p_match}
\end{figure*}

We must still construct a bank of waveforms to cover the  four-dimensional parameter space of masses, the effective
aligned $\chi_{\mathrm{eff}}$ and precessing $\chi_{P}$ components of the spins.  The amplitude and phase evolution
of a single harmonic does not carry the tell-tale amplitude and phase modulation caused by precession, but
does have a different phase evolution due to precession \cite{Lundgren:2013jla, Hannam:2013oca}.  Since the phase
evolution of each precessing harmonic is degenerate with a non-precessing waveform with different mass-ratio
or effective spin, the bank of templates will essentially be a bank of non-precessing waveforms.  This
may allow us to reduce the size of the template bank.

The k=0 harmonic of the precessing waveform has an additional phase (see Eq.~(\ref{eq:prec_phase}))
of,
\begin{equation}\label{eq:dphi_0}
	\delta \phi_{0}(t) = \int_{t_{o}}^{t} \frac{2b^{2}}{1 + b^{2}}  \, \dot{\alpha} \, dt' \, .
\end{equation}
For systems in which orbital angular momentum dominates over spin angular momentum, 
the precession frequency is inversely proportional to orbital frequency, $\Omega_{P} = \dot{\alpha} \propto f^{-1}$
\cite{Apostolatos:1994mx, Buonanno:2002fy, Brown:2012gs}.  This is the same frequency dependence as the 
1PN contribution to the waveform, whose amplitude depends upon the mass ratio.
Consequently, it is reasonable to expect that the precession-induced phase will be indistinguishable from 
a systematic offset in the binary mass ratio, or the effective spin \cite{Hannam:2013uu}.  Similarly,
the $k=1$ harmonic has essentially the same amplitude evolution as the non-precessing waveform,
but with a phase difference of,
\begin{equation}\label{eq:dphi_1}
	\delta \phi_{1}(t) = - \int_{t_{o}}^{t} \frac{1 - b^{2}}{1 + b^{2}}  \, \dot{\alpha} \, dt' \, ,
\end{equation}
which will also, in many cases, be degenerate with a change in the mass ratio or aligned spin.

In Figure \ref{fig:chi_eff_chi_p_match}, we investigate the degeneracy in the spin ($\chi_{\mathrm{eff}}$--$\chi_{P}$)
space of the two leading precession harmonics.  We consider a system with masses, $M = 40 M_{\odot}$ and $q = 4$,
and spins $\chi_{\mathrm{eff}} = 0$, $\chi_{p} = 0.6$ and investigate how the two waveform harmonics vary as the
spins are changed.  The figure shows the match --- the overlap maximized over time-offsets --- between our fiducial
waveform and one with the same masses but different spins.  
For both harmonics, there is a band in the $\chi_{\mathrm{eff}}$--$\chi_{P}$ plane where the is mismatch is small ---
the different phase evolution of each harmonic caused by varying $\chi_{P}$
can be offset by a suitable change in $\chi_{\mathrm{eff}}$.  The relation is approximately quadratic, 
$\Delta \chi_{\mathrm{eff}} \propto (\Delta \chi_{P})^{2}$, which is to be expected.  
Recall, from Eq.~(\ref{eq:dphi_0}), that  change in phase due to precession
is quadratic in $b$, and therefore also in $\chi_{P}$ at least for small values of $b$.  Meanwhile the
phasing of the waveform varies, at leading order, linearly with $\chi_{\mathrm{eff}}$.

This degeneracy in the $\chi_{\mathrm{eff}}$--$\chi_{P}$ plane suggests that a single template waveform could be
used to search over an extended region corresponding, for example, 
to the region of mismatch $< 0.03$ in Figure \ref{fig:chi_eff_chi_p_match}.  
However, this will only work if the degenerate region for the $k = 0$ and $k=1$ harmonics is the same.  It is clear from 
Equations (\ref{eq:dphi_0}) and (\ref{eq:dphi_1}) and Figure~\ref{fig:chi_eff_chi_p_match} that they are not identical.
Nonetheless,%
\footnote{Strictly, when doing this comparison, we must use the same time offset for the two harmonics, whereas
the figure allows for an independent maximization of the time delay for each harmonic.  Fixing a single time
delay does slightly decrease the matches, but not significantly enough to change the conclusions.}
for the example we have considered, the two degenerate regions are similar, and along the line that traces the 
mid-point between  best fit values of $\chi_{\mathrm{eff}}$ for the two harmonics, both harmonics have a match above 
0.97 with the initial point.  Thus, to an accuracy appropriate for generating a template bank, we can use the two harmonics 
from a single waveform to cover a band in the $\chi_{\mathrm{eff}}$--$\chi_{P}$ plane which spans all values of $\chi_{P}$.
This effectively reduces the dimensionality of the parameter space to three dimensions: mass, 
mass ratio and one spin parameter.  

Our proposal to develop a precessing search is as follows.  First, generate a bank of templates to cover the
space of non-precessing binaries.  At each $M$, $q$, $\chi_{\mathrm{eff}}$ point in the template bank,
construct the two-harmonic waveform for a fixed value of $\chi_{P}$.
Then, filter the data against the two harmonics and calculate the two-harmonic SNR, as defined in
Eq.~(\ref{eq:rho_2harm}) to identify candidate events in a single detector.
It will be necessary to extend the existing $\chi^{2}$ signal consistency test \cite{Allen:2004gu} to each harmonic, 
taking into account the presence of the other harmonics, to
reduce the impact of non-stationarity in the data.  Next, perform coincidence between detectors by requiring
a signal in the same template at the same time, up to the allowed time delays based upon speed of propagation.
For a non-precessing signal observed
in two detectors, the relative amplitude and phase of the SNR in each detector can take any value, even
though some are astrophysically more likely \cite{Nitz:2017svb} (and this can be used to increase search sensitivity).
However, for the two-harmonic waveform not every
signal observed in two detectors will be compatible with an astrophysical source.  This can be seen through simple
parameter counting: there are ten measured quantities (two complex amplitudes and a time of arrival in each detector),
which depend upon eight parameters, the five orientation parameters ($d_{L}$, $\theta_{\rm JN}$, $\psi$,
$\phi_{o}$, $\alpha_{o}$), sky location and merger time.  An additional coincidence test to check for consistency
between parameters will likely be necessary to reduce the search background.  A similar problem arises already in
extending the amplitude and phase consistency of \cite{Nitz:2017svb} to three or more detectors and methods 
developed for that purpose may be helpful for the precessing search.

We can estimate the likely sensitivity improvement from a precessing search, as we have briefly discussed in 
Section \ref{sec:validity}.  A non-precessing search
will typically find the dominant harmonic of the waveform.  Thus, for signals where two harmonics provide a
significant contribution, a search based on the two-harmonic waveform has the potential
to out-perform the non-precessing search.
The two-harmonic waveform has four degrees of freedom, encoded in $\mathcal{A}_{0}$ and $\mathcal{A}_{1}$,
compared to two for the non-precessing search.  Thus, the noise background is higher for the
two-harmonic search and, based upon a comparison of the tails of the $\chi^{2}$ distribution with 2 and 4
degrees of freedom, an increase of around 5\% in SNR is required to obtain the same false alarm rate 
(see e.g., Ref.~\cite{Harry:2016ijz} for a discussion of this issue).  
Thus, a signal will be observed as more significant in the two-harmonic search than a non-precessing search
if the SNR can be increased by 5\% or more.  Fig.~\ref{fig:pop_overlap} shows that this occurs for 
$\overline{b} \gtrsim 0.15$, and for binaries with $\overline{b}$ above this value the two-harmonic search has the
potential to outperform a non-precessing search.  We note, however, that a given template will cover a range of
spin values and consequently a range of $\overline{b}$, so it may be more appropriate to deploy the
two-harmonic search for templates with an \textit{average} of $\overline{b}$ which is greater than $0.15$.

Another challenge of searches for precessing systems is the associated computational cost \cite{Harry:2016ijz}, which
can be prohibitive.  The maximum computational cost for the two-harmonic search would be double that of a
comparable non-precessing search: it becomes necessary to filter both the $k=0$ and 1 harmonics, and computational time
is dominated by this matched filtering.  However, since both the $k=0$ and $k=1$ harmonics are essentially non-precessing 
waveforms, there may be waveforms associated with the $k=1$ harmonics are \textit{already} in the set of $k=0$ waveforms, 
but associated with different parameters.  If so, this could further reduce the computational cost.

\section{Observability of precession}
\label{sec:obs_prec}

\begin{figure*}[t]
	\begin{center}
	\includegraphics[width=\textwidth]{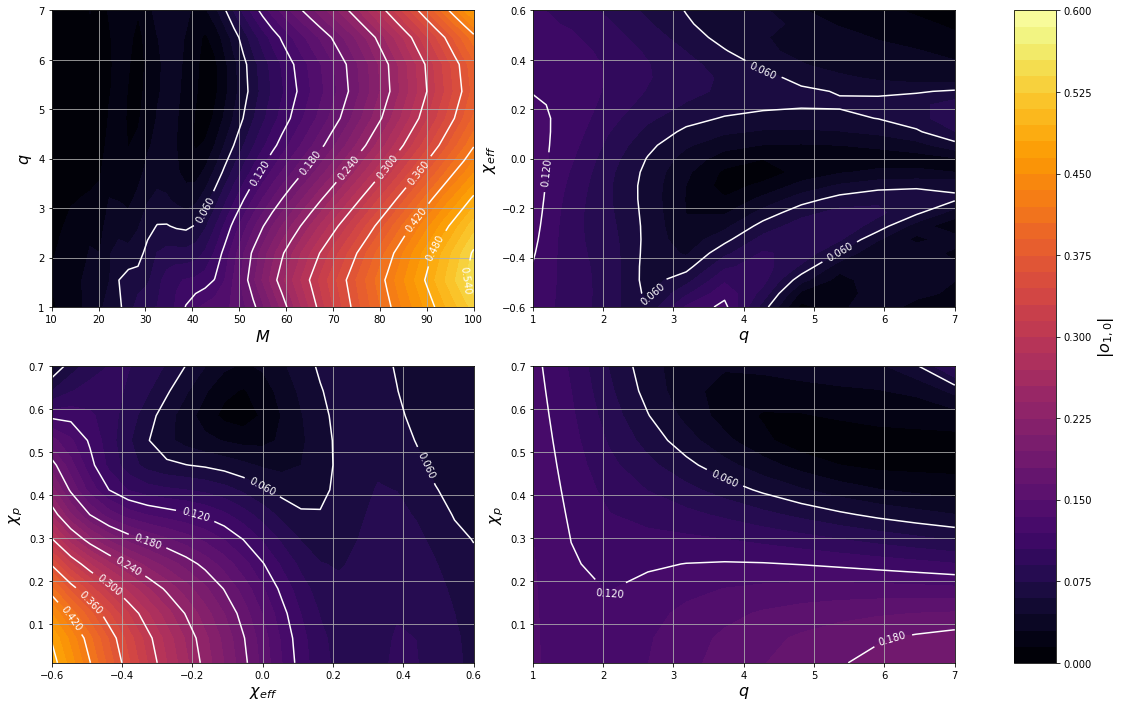}
	\end{center}
	\caption{
	The overlap $O(h_{0}, h_{1})$ between the $k=0$ and $k=1$ harmonics across two-dimensional slices in the 
	parameter space of  total mass, mass ratio, $\chi_{\mathrm{eff}}$ and $\chi_{p}$.
	In each plot, two of the parameters are varied while the other two are fixed to their fiducial values of
	$M = 40 M_{\odot}$, $q = 4$, $\chi_{\mathrm{eff}} = 0$, $\chi_{p} = 0.6$ .	}
	\label{fig:overlap_parameter_space}
\end{figure*}

The two-harmonic approximation allows us to easily identify regions of the binary merger parameter space for 
which precession will leave an observable imprint on the waveform.  Since the amplitude and phase evolution 
of a single harmonic is generally consistent with that of a non-precessing waveform (see above and 
\cite{Lundgren:2013jla, o2020semianalytic}), it is only when two harmonics
can be observed that we are able to clearly identify precession in the system.  We
are therefore interested in deriving an expression for the \textit{precession SNR}, $\rho_{p}$, defined as the
SNR in the second most significant harmonic, and determining when it will be observable.  
If the two harmonics $h^{0}$ and $h^{1}$ in Eq.~(\ref{eq:2harm}) are orthogonal, then the precession SNR is simply,
\begin{eqnarray}
	\rho_{p} &=& \mathrm{min}(|\mathcal{A}_{0} h^{0}|, |\mathcal{A}_{1} h^{1}| ), \nonumber \\
	&=&  \rho_{\rm 2harm} \, \left(\frac{ \mathrm{min}(1, |\zeta| )}{ \sqrt{1 + |\zeta|^{2}}}\right),
	\label{eq:rho_prec_orth}
\end{eqnarray}
where $\zeta$, defined in Eq.~(\ref{eq:zeta}), gives the ratio of the SNR in the $k=1$ and $k=0$ harmonics and
$\rho_{\mathrm{2harm}}$ is the total SNR in the two-harmonic waveform.

Let us briefly examine where in parameter space the two harmonics are
close to orthogonal.  Where there are sufficient precession cycles we expect the two
harmonics, $h^{0}$ and $h^{1}$, will be close to orthogonal, and the overlap to be
close to zero \cite{Lundgren:2013jla}.  The overlap between the two harmonics for various two-dimensional slices through 
the parameter space is shown in Fig.~\ref{fig:overlap_parameter_space}.
At higher masses, where the binary completes one, or fewer, precession cycles
in the detector's sensitive band, there is a larger overlap between the harmonics.  At negative
$\chi_{\mathrm{eff}}$ and minimal $\chi_{p}$, the overlap is also significant.  However, providing the
mass of the system is below $50 M_{\odot}$, for the much of the parameter
space the overlap is less than $0.1$ and simple expression in Eq.~(\ref{eq:rho_prec_orth}) will be
applicable.

Taking into account the overlap between harmonics, the total power in the two-harmonic waveform is,
\begin{equation}
	\rho^{2}_{\rm 2harm} = |\mathcal{A}_{0} h^{0} |^{2} \left( 1 + 2 \mathrm{Re} [\zeta \, o_{1,0}] + |\zeta|^{2} \right) \, .
\end{equation}
where $o_{1,0}$ is complex overlap between the two harmonics:
\begin{equation}\label{eq:complex_overlap}
	o_{1,0} = \frac{(h^{1} | h^{0}) + i (h^{1} | i h^{0} )}{| h^{1} | | h^{0} |} \, .
\end{equation}
We can project the SNR onto directions parallel and perpendicular to the $h^{0}$ waveform to obtain the SNR in
these two directions as,
\begin{eqnarray}
	\rho_{0}^{2} &=& |\mathcal{A}_{0} h^{0} |^{2} \left( 1 + 2 \mathrm{Re} [\zeta \, o_{1,0}] + |\zeta \, o_{1,0} |^{2} \right), \nonumber \\
	\rho_{\perp, 0}^{2} &=& |\mathcal{A}_{0} h^{0} |^{2} |\zeta|^{2} \left(1 - | o_{1,0} |^{2} \right) \, .
\end{eqnarray}
Similarly, the power parallel to and perpendicular to the $k=1$ harmonic is,
\begin{eqnarray}
	\rho_{1}^{2} &=& |\mathcal{A}_{0} h^{0} |^{2} \left( |o_{1,0} |^{2} + 2 \mathrm{Re} [\zeta \, o_{1,0}] + |\zeta |^{2} \right), \nonumber \\
	\rho_{\perp, 1}^{2} &=& | \mathcal{A}_{0} h^{0} |^{2} \left(1 - | o_{1,0} |^{2} \right) \, .
\end{eqnarray}

The precession SNR is defined as the power orthogonal to the dominant harmonic,%
\footnote{In exceptional circumstances, where the overlap is large and $\zeta o_{1,0}$ is close to $-1$, there can be 
more power in $\rho_{\perp, i}$ than $\rho_{i}$.  In such cases, it is natural to use $\rho_{i}$ to determine
if precession is present, although this is not ideal as $\rho_{\perp, i}$ need not resemble a non-precessing waveform.}
\begin{eqnarray}
	\rho_{p}
	&:=& \mathrm{min} (\rho_{\perp, 0} , \rho_{\perp, 1}), \\
	&=& \rho_{\rm 2harm} \, \mathrm{min} (1, |\zeta|)
	\left( \frac{1 - |o_{1,0}|^{2}}{1 + 2 \mathrm{Re}[\zeta\,  o_{1,0}] + |\zeta|^{2} }\right)^{\frac{1}{2}} \, . \nonumber 
\end{eqnarray}
As expected, the precession SNR scales with the total SNR of the signal, so that precession will be
more easily observed for louder events.  If there is significant degeneracy between the harmonics,
the numerator will be reduced, making the observation of precession more difficult.  Finally,
in the limit that $o_{1,0} \rightarrow 0$, the expression simplifies to the one
given earlier for orthogonal harmonics in (\ref{eq:rho_prec_orth}), as expected.

What value of $\rho_{p}$ will be required to observe precession?  This will happen if the evidence for a signal with
$\chi_{p} \neq 0$ in the data is greater than that for a non-precessing source.  This can be evaluated through Bayesian
model selection, by considering the Bayes factor between the hypotheses.  However, such a calculation requires a full
exploration of the parameter space.  We can, instead, obtain an approximate answer by considering the maximum
likelihood.  Since the two-harmonic waveform is more general than the non-precessing waveform, it will always give
a larger maximum likelihood \textit{even in the absence of precession} do to its ability to fit the detector noise.  Thus,
we are interested in examining the expected increase in SNR due to the inclusion of the second harmonic, in the
absence of any power in it.  

The two-harmonic SNR can be written as
\begin{equation}
	\rho_{\mathrm{2harm}}^{2} = \rho_{np}^{2} + \rho_{p}^{2} \, .
\end{equation}
where $\rho_{np}$ is the non-precessing SNR or, equivalently, the SNR in the dominant harmonic.
In the absence of precession, $\rho_{p}$ will be $\chi^{2}$ distributed with 2 degrees of freedom, as we are able to freely 
maximize over the amplitude and phase of the two harmonics independently \cite{Allen:2005fk, Babak:2012zx}.  
Consequently, in 90\% of cases, noise
alone will give a value of $\rho_{p} < 2.1$.  Therefore, as a simple criterion, we require that,
\begin{equation}
	\rho_{p} \ge 2.1,
\end{equation}
for precession to be observable.  In Ref.~\cite{Green:2019_prec_pe} we use this definition to investigate in
detail the observability of precession over the parameter space.

\section{Discussion}
\label{sec:disc}

We have presented a new, intuitive way to understand the observability of precession in
GW observations.  By keeping only the leading precession term, we have derived a
precession SNR and argued that this can be used to determine when precession will be
observable. Before discussing applications we point out the main limitations of this analysis. As
is clear from the formulation, this analysis works best for binaries where $b = \tan(\beta/2)$ is small.
This typically corresponds to situations where the masses are comparable, the precessing spin is small
and any aligned component of the spin is aligned (rather than anti-aligned) with the orbital angular
momentum. We have shown above that this assumption is valid for a reasonable population.

We now point to several advantages and applications of this formulation:
\newline
First, it gives new understanding of the observability of precession, and also of the origin of
precession as the beating of two waveform components with slightly differing frequencies 
(also discussed in \cite{Lundgren:2013jla}).
It is difficult to identify the presence of precession in a GW observation directly from $\chi_P$, since the prior astrophysical 
expectation disfavours $\chi_P = 0$.  While the deviation from the prior can be determined through the 
Bayes factor, the results in this paper suggest that the precession SNR $\rho_P$ could provide a direct measure of whether 
precession has been measured in a signal. The potential applications of $\rho_P$ are discussed in the companion 
paper~\cite{fairhurst2019will}, and will be investigated in more detail in Ref.~\cite{Green:2019_prec_pe}, where we probe the 
measurability of precession across the gravitational wave parameter space.

There exist a number of detailed population analyses which extract the features of the underlying population of gravitational
waves from the set of observed gravitational wave events, for example~\cite{Talbot:2017yur, Talbot:2019okv, Wysocki:2018mpo, 
LIGOScientific:2018jsj}. 
These typically use the full posterior distributions recovered from the gravitational wave signal~\cite{Veitch:2014wba, Ashton:2018jfp} 
to infer the population and, as
such, naturally account for precession effects in the observed signals when inferring the black hole mass and spin populations.  
Nonetheless, there have been a number of studies performed which investigate the population properties using a subset of
the recovered parameters, see e.g.~\cite{Farr:2017gtv, Farr:2017uvj, Fishbach:2017zga, Fishbach:2018edt, Tiwari:2018qch, LIGOScientific:2018jsj}, and have been 
successfully used to infer interesting properties of the mass and spin distributions.  The majority of these studies have restricted 
attention to the aligned components of the spins.  The precession SNR provides a straightforward method to determine the 
significance of precession, and provides away to probe observability of precession in populations of binaries. In using this method 
we have been able to derive constraints on the preferred spin distribution including precession effects~\cite{fairhurst2019will}.

Both of the applications highlighted above are currently possible using other more sophisticated but computationally expensive
methods such as Bayesian model comparison. This is, of course, a more general method that makes fewer
assumptions than we do in computing $\rho_p$, however the computational costs associated with calculating the marginal 
likelihood over multiple, e.g.~precessing and non-precessing, models per binary are not feasible for a large number of binaries. 
For example the analysis in \cite{fairhurst2019will} involved calculating  $\rho_p$  for 1 million binaries, and 
computing the Bayes factor for 1 million binaries would certainly not be practical.  Similar, lightweight analyses, could also be 
developed using the formalism introduced in, e.g. \cite{Brown:2012gs}, and if this is done,it would be interesting to compare them
with the results from the two harmonic analysis.

Finally, we have outlined a method by which the two-harmonic approximation could be used to develop a search for precessing
binaries. We have shown that in principle that this approach could result in a significant increase in sensitivity 
without the computational overheads associated with other precessing search methods.
In addition, the formalism should provide a way to  identify the parts of parameter space where a precessing search is
likely to increase sensitivity. We plan a detailed investigation into the feasibility of a precessing search based 
upon the  two-harmonic approximation in future work.

\section*{Acknowledgements}

We would like to thank Eleanor Hamilton, Ian Harry, Andrew Lundgren, Frank Ohme, Francesco Pannarale, 
Vivien Raymond, Patrick Sutton and Vaibhav Tiwari for discussions. 
This work was supported by Science and Technology Facilities Council (STFC) grant ST/L000962/1, European
Research Council Consolidator Grant 647839 and the National Science Foundation under Grant No. NSF PHY-1748958.
We are grateful for computational resources provided by Cardiff University, and funded by an
STFC grant supporting UK Involvement in the Operation of Advanced LIGO.

\appendix*

\section*{Appendix: Derivation using spin-weighted spherical harmonics}

In this appendix, we provide an alternative derivation of the power series decomposition of the precessing waveform,
given in Section \ref{sec:waveform}, based upon the spin-weighted spherical harmonic decomposition of the waveform 
\cite{Thorne:1980ru} and its application to precession as described in \cite{Hannam:2013oca, Boyle:2011gg}.  Specifically,
we wish to obtain the result in Eq.~(\ref{eq:h_prec}).  Throughout, we follow the notation used in \cite{Khan:2019kot}.

The gravitational waveform emitted during a binary merger, 
\begin{equation}
h := h_{+} - i h_{\times} 
\end{equation}
can naturally be decomposed into a set of spin-weighted spherical harmonics as
\begin{equation}
h(t, \vec{\lambda}, \theta, \alpha_{o}) = 
\sum_{\ell \ge 2} \sum_{-\ell \le m \le \ell} h_{\ell, m}(t, \vec{\lambda}) {}^{-2}Y_{\ell, m}(\theta, \phi)
\end{equation}
where $\theta$ and $\phi$ give the orientation of the observer relative to a co-ordinate system used to identify
the spherical harmonics, $\vec{\lambda}$ encodes the physical parameters of the system (masses, spins, etc) and $t$ is the time.

The multipoles for a precessing system are approximated by ``twisting up'' \cite{Hannam:2013oca, Boyle:2011gg}
the multipoles of the non-precessing counterpart based upon the orientation of the orbital angular momentum given by
the opening angle $\beta$, precession angle $\alpha$ and the third Euler angle $\epsilon$ defined via
\begin{equation}
\dot{\epsilon} = \dot{\alpha} \cos \beta \, .
\end{equation}
Then, the precessing multipoles are given by
\begin{equation}
h^{\mathrm{prec}}_{\ell, m}(t) = \sum_{-\ell \le n \le \ell} h^{\mathrm{NP}}_{\ell, n} D^{\ell}_{n, m}(\alpha(t), \beta(t), \epsilon(t))
\end{equation}
where the Wigner D-matrix is 
\begin{equation}
D^{\ell}_{n, m}(\alpha, \beta, \epsilon) = e^{i m \alpha} d^{\ell}_{n, m}(-\beta) e^{-i n \epsilon}
\end{equation}
and the Wigner d-matrix given, for example, in \cite{Brown:2007jx}.

Combining these decompositions gives the waveform for a precessing binary as
\begin{equation}\label{eq:prec_modes}
h = \sum_{\ell, m, n} {}^{-2}Y_{\ell, m}(\theta, \phi) D^{\ell}_{n, m}(\alpha, \beta, \epsilon) h_{\ell, n}(t, \vec{\lambda}) \, .
\end{equation}
In performing the twisting, it's natural that the precessing waveform is described in a coordinate system aligned
with the orbital angular momentum, so that $\theta = \theta_{\mathrm{JN}}$. Furthermore, the orientation of the $x$-axis
will be specified relative to the (initial) precession phase so that $\phi = - \alpha_{o}$.

In this work, we restrict attention to the case where the non-precessing model contains only the $\ell = 2$ and $n = \pm 2$
modes, and require symmetry in gravitational wave emission above and below the plane of the binary so that 
$h_{\ell, n} = (-1)^{\ell} h^{\star}_{\ell, -n}$.  This eliminates the sum over $\ell$ and $m$ from Eq.~(\ref{eq:prec_modes}).
Furthermore, we can expand the spherical harmonics using
\begin{equation}
{}^{-2}Y_{2, m}(\theta_{\mathrm{JN}}, -\alpha_o) = 
\sqrt{\frac{5}{4\pi}} d^{2}_{m, 2}(\theta_{\mathrm{JN}}) e^{- im \alpha_{o}}
\end{equation}
to obtain
\begin{eqnarray}\label{eq:prec22_modes}
h^{\mathrm{prec}} &=& \sum_{-2 \le m \le 2} \sqrt{\frac{5}{4\pi}} d^{2}_{2, m} (\theta_{\mathrm{JN}}) e^{i m (\alpha - \alpha_{o})}
\times \\
&& \left[ h^{\mathrm{NP}}_{22} d^{2}_{2, m}(- \beta) e^{- 2 i \epsilon} 
+ (h^{\mathrm{NP}}_{22})^{\star} d^{2}_{-2, m}(- \beta) e^{2 i \epsilon} \right] \nonumber
\end{eqnarray}

We now wish to re-write the above to show that the waveform can be decomposed in modes whose amplitudes
form a power series in $b = \tan(\beta/2)$.  To do so, we note that the Wigner d-matrices can be evaluated as powers of 
$\sin(\beta/2)$ and $\cos(\beta/2)$, so that if we are able to group terms with the same indices we will arrive at the desired 
expression.  We do this by using the d-matrix identities:
\begin{equation}
	d^{\ell}_{n, m} = (-1)^{m-n} d^{\ell}_{m, n} = (-1)^{m - n} d^{\ell}_{-n, -m}
\end{equation}
and relabelling the dummy index $m \rightarrow -m$ in the second term of Eq.~(\ref{eq:prec22_modes}) to obtain:
\begin{eqnarray}
h^{\mathrm{prec}} &=& \sum_{-2 \le m \le 2} \sqrt{\frac{5}{4\pi}} d^{2}_{2, m} (- \beta) \times \\
&& \left[ (-1)^{m} d^{2}_{2, m}(\theta) \left( h^{\mathrm{NP}}_{22}(t) e^{- 2 i \epsilon} e^{i m (\alpha - \alpha_{o})} \right) \right. \nonumber \\
&& \left. \; + d^{2}_{2, -m}(\theta) \left( h^{\mathrm{NP}}_{22}(t) e^{- 2 i \epsilon} e^{i m (\alpha - \alpha_{o})} \right)^{\star}  \right] \nonumber
\end{eqnarray}

Finally, we can evaluate the Wigner d-matrices as
\begin{eqnarray}
	d^{2}_{2,  m}(-\beta) 
	&:=& C_{m} \cos^{2+m}(\beta/2) \sin^{2-m}(\beta/2) \nonumber \\
	&=& \frac{C_{m} b^{2-m}}{( 1 + b^{2})^{2}} 
\end{eqnarray}
where $C_{\pm2} = 1$, $C_{\pm 1} = 2$, $C_{0} = \sqrt{6}$ and, as before, $b = \tan(\beta/2)$.  Similarly, we introduce
$\tau = \tan \theta_{JN}/2$, and evaluate the d-matrices for the angle $\theta_{\mathrm{JN}}$.  This gives
\begin{eqnarray}
h^{\mathrm{prec}} &=& \sum_{-2 \le m \le 2} \sqrt{\frac{5}{4\pi}} \frac{(C_m)^{2} b^{2-m}}{( 1 + b^{2})^{2}} \times \\
&& \left[ \frac{\tau^{2-m}}{(1 + \tau^{2})^{2}} \left(h^{\mathrm{NP}}_{22}(t) e^{- 2 i \epsilon} e^{i m (\alpha - \alpha_{o})}\right) \right. \nonumber \\
&& \left. \; + \frac{(-\tau)^{2+m}}{(1 + \tau^{2})^{2}} \left(h^{\mathrm{NP}}_{22}(t) e^{- 2 i \epsilon} e^{i m (\alpha - \alpha_{o})}
\right)^{\star} \right] \nonumber
\end{eqnarray}

This is close to the desired form and, in particular, we have obtained an decomposition where the relative strength of
each mode is decreased by a factor of $b$.  To obtain an expression comparable to Eq.~(\ref{eq:h_prec}) we must 
evaluate the waveform observed at a detector with response $F_{+}$ and $F_{\times}$ to the two gravitational
polarizations.
\begin{eqnarray}
	h(t) &=& 
	\mathrm{Re} \left[ (F_{+} + i F_{\times}) h^{\mathrm{prec}} \right]  \\
	&=& \mathrm{Re} \left[ \left(\sqrt{\frac{5}{4\pi}} \frac{(h^{\mathrm{NP}}_{22})^{\star} e^{2 i (\epsilon + \alpha - \alpha_{o})}}
	{(1 + b^{2})^{2}} 
	 \right) \right.
	\nonumber \\
	&& \sum_{m = -2}^{2}  \frac{(C_m)^{2}}{(1 + \tau^{2})^{2}} ( be^{-i(\alpha - \alpha_{o})} )^{2-m} 
	\nonumber \\
	&& \left. \quad \left( F_{+} [\tau^{2-m} + (-\tau)^{2+m}] - i F_{\times} [\tau^{2-m} - (-\tau)^{2+m}] \right) \;  \right] \nonumber
\end{eqnarray}
Then, to finally equate this with the desired expression, we must make the identification
\begin{equation}
 \sqrt{\frac{5}{4\pi}} (h^{\mathrm{NP}}_{22})^{\star} e^{2 i \epsilon} = 
 \frac{d_{o}}{d_L} A_{o}(t) e^{2 i \Phi_{S}} \, ,
\end{equation}
where $\Phi_{S}$ is defined in Eq.~(\ref{eq:phi_s}).  Thus the amplitude of the waveform, $A_{o}(t)$ is the same as the scaled 
$22$ mode while the phase of the 22 mode is the (negative) of the orbital phase.  Furthermore, it is straightforward
to show that the $\mathcal{A}^{+,\times}_{k}$ coefficients are given by
\begin{align}
\mathcal{A}_{(2-m)}^+ &= \frac{d_{o}}{d_L} (C_m)^{2} \left( \frac{\tau^{2-m} + (-\tau)^{2+m}}{(1 + \tau^{2})^2} \right), \nonumber\\
\mathcal{A}_{(2-m)}^{\times} &= \frac{d_{o}}{d_L} (C_m)^{2} \left( \frac{\tau^{2-m} - (-\tau)^{2+m}}{(1 + \tau^{2})^2} \right). \nonumber\\
\end{align}
Substituting these identifications, we obtain the desired expression for the waveform observed at a detector,
\begin{eqnarray}
	h(t) &=& \mathrm{Re}
	\left[ \left( \frac{A_{o}(t) e^{2 i (\Phi_{S} + \alpha)}}{(1 + b^{2})^{2}} \right)
	\right. \nonumber \\
	&& \qquad \left.
	\sum_{k=0}^{4}  (b e^{-i\alpha})^{k}
	(F_{+} \mathcal{A}_{k}^{+} - i F_{\times} \mathcal{A}_{k}^{\times}) \right] \, .
\end{eqnarray}

\bibliographystyle{../unsrt_LIGO}
\bibliography{../Paper_References}

\end{document}